\documentclass[sigconf]{acmart}
\AtBeginDocument{%
  }

\setcopyright{acmlicensed}
\copyrightyear{2026}
\acmYear{2026}
\acmDOI{XXXXXXX.XXXXXXX}

\acmConference[ICSE '26]{International Conference on Software Engineering}{April 12-18,
  2026}{Rio de Janeiro, Brazil}
\acmISBN{978-1-4503-XXXX-X/2018/06}
\usepackage{enumitem}
\usepackage{booktabs}
\usepackage{multirow}
\usepackage{colortbl}
\usepackage{xcolor}
\usepackage{bigstrut}
\usepackage{array}
\usepackage{amsmath}
\usepackage{makecell}

\usepackage[framemethod=tikz]{mdframed}
\newcounter{finding}
\newcommand{\finding}[1]{\refstepcounter{finding}
  \vspace{2.3mm}
 \begin{mdframed}[linecolor=gray,roundcorner=12pt,backgroundcolor=gray!15,linewidth=3pt,innerleftmargin=2pt, leftmargin=0cm,rightmargin=0cm,topline=false,bottomline=false,rightline = false]
  \textbf{Ans. to RQ\arabic{finding}:} #1
 \end{mdframed}
 \vspace{2.3mm}
}

\begin{document}

%%
%% The ``title'' command has an optional parameter,
%% allowing the author to define a ``short title'' to be used in page headers.
\title{\color{black}Fairness Is Not Just Ethical: Performance Trade-Off via Data Correlation Tuning to Mitigate Bias in ML Software\color{black}}

%%
%% The ``author'' command and its associated commands are used to define
%% ``authornote'' and ``authornotemark'' commands

%% The ``author'' command and its associated commands are used to define
%% ``authornote'' and ``authornotemark'' commands
\author{Ying Xiao}
\email{ying.1.xiao@kcl.ac.uk}
\affiliation{%
  \institution{King's College London}
  % \city{London}
  \country{United Kingdom}
}

\author{Shangwen Wang}
\affiliation{%
  \institution{National University of Defense Technology}
  % \city{London}
  \country{China}
}

\author{Sicen Liu}
\affiliation{%
  \institution{Southern University of Science and Technology}
  % \city{London}
  \country{China}
}

\author{Dingyuan Xue}
\affiliation{%
  \institution{Southern University of Science and Technology}
  % \city{London}
  \country{China}
}

\author{Xian Zhan}
\affiliation{%
  \institution{The Hong Kong Polytechnic University}
  % \city{London}
  \country{China}
}

\author{Yepang Liu}
\affiliation{%
  \institution{Southern University of Science and Technology}
  % \city{London}
  \country{China}
}

\author{Jie M. Zhang}
\authornote{Corresponding author.}
% \email{jie.zhang@kcl.ac.uk}
\affiliation{%
  \institution{King's College London}
  % \city{London}
  \country{United Kingdom}
}

\renewcommand{\shortauthors}{Xiao et al.}
%%
%% Article type: Research, Review, Discussion, Invited or position
\acmArticleType{Research}
%%
%% Links to code and data
\acmCodeLink{https://github.com/borisveytsman/acmart}
\acmDataLink{htps://zenodo.org/link}
%%
%% Authors' contribution
\acmContributions{BT and GKMT designed the study; LT, VB, and AP
  conducted the experiments, BR, HC, CP and JS analyzed the results,
  JPK developed analytical predictions, all authors participated in
  writing the manuscript.}

\begin{abstract}

% Regarding software fairness solely as an ethical imperative can underestimate the practical significance of software fairness research. In reality, software fairness fundamentally represents a software quality issue, arising directly from performance disparities across user groups. Bias mitigation often provides an added benefit: improved predictive performance for unprivileged groups. This shows that bias mitigation techniques can actively adjust predictive performance across groups, significantly enhancing software system performance in out-of-distribution scenarios and geographic transferability. However, existing bias mitigation methods face a dilemma: pre-processing methods are more widely applicable than post-processing methods, but typically lag behind in effectiveness. To address this challenge, we propose Correlation Tuning (CoT), a novel pre-processing approach that mitigates bias by adjusting data correlations. We introduce the Phi-coefficient to measure the correlation between sensitive attributes and labels and employ it, together with multi-objective optimization, to guide fairness improvements. Extensive evaluation demonstrates that CoT increases the true positive rate for unprivileged groups by an average of 20.0\% and reduces key fairness biases (SPD, AOD, EOD) by over 50\% on average, surpassing the state-of-the-art by 5 and 8 percentage points in single-attribute and multi-attribute scenarios, respectively. We release our results, source code, and fine-tuned models to facilitate and advance future software fairness research.

Traditional software fairness research typically emphasizes ethical and social imperatives, neglecting that fairness fundamentally represents a core software quality issue arising directly from performance disparities across sensitive user groups. Recognizing fairness explicitly as a software quality dimension yields practical benefits beyond ethical considerations, notably improved predictive performance for unprivileged groups, enhanced out-of-distribution generalization, and increased geographic transferability in real-world deployments. Nevertheless, existing bias mitigation methods face a critical dilemma: while pre-processing methods offer broad applicability across model types, they generally fall short in effectiveness compared to post-processing techniques. To overcome this challenge, we propose Correlation Tuning (CoT), a novel pre-processing approach designed to mitigate bias by adjusting data correlations. Specifically, CoT introduces the Phi-coefficient, an intuitive correlation measure, to systematically quantify correlation between sensitive attributes and labels, and employs multi-objective optimization to address the proxy biases. Extensive evaluations demonstrate that CoT increases the true positive rate of unprivileged groups by an average of \textbf{17.5\%} and reduces three key bias metrics, including \color{black}statistical parity difference (SPD), average odds difference (AOD), and equal opportunity difference (EOD)\color{black}, by more than \textbf{50\%} on average. CoT outperforms state-of-the-art methods by three and ten percentage points in single attribute and multiple attributes scenarios, respectively.  We will publicly release our experimental results and source code to facilitate future research.

\end{abstract}



%%
%% Sometimes the addresses are too long to fit on the page.  In this
%% case uncomment the lines below and fill them accodingly.
%%
%% \authorsaddresses{Corresponding author: Ben Trovato,
%% \href{mailto:trovato@corporation.com}{trovato@corporation.com};
%% Institute for Clarity in Documentation, P.O. Box 1212, Dublin,
%% Ohio, USA, 43017-6221}

\maketitle

\section{Introduction}
\label{sec:intro}

Empowered by artificial intelligence (AI) techniques such as machine learning (ML) and deep learning (DL), software systems have become increasingly intelligent and widely deployed in critical decision-making scenarios, including justice \cite{misc_compas_dataset, propublicaAnalyzedCOMPAS, chen2022maat}, healthcare \cite{ahrqMedicalExpenditure-mep, ahrqMedicalExpenditure-mep2, xiao2024mirrorfair}, and finance \cite{misc_south_german_credit_522, misc_adult_2, chakraborty2021bias}. However, these software systems may produce biased predictions against groups or individuals with specific sensitive attributes, causing significant concern about software fairness \cite{chakraborty2021bias, chen2022fairness} and potentially violating anti-discrimination laws \cite{zhang2021ignorance}. In the Software Engineering (SE) community, software fairness is generally considered an ethical issue, and developing responsible, fair software is recognized as an essential ethical responsibility for software engineers \cite{chen2022maat, chen2024fairness}. In fact, fairness metrics such as Equal Opportunity Difference quantify the performance disparity across groups \cite{chakraborty2021bias, biswas2023fairify}. Meanwhile, performance and consistency are critical software quality attributes explicitly outlined by established quality frameworks, such as McCall’s Quality Model \cite{mccall1977factors}, Boehm’s Quality Model \cite{boehm1976quantitative}, and the FURPS Quality Model \cite{grady1992practical, al2010quality}. Thus, software fairness represents \textbf{not only a social and ethical concern} but also a \textbf{fundamental software quality issue} arising from software performance disparities.

Although substantial progress has been made in the SE community to improve software fairness, one conceptual gap and one significant challenge hinder further advancement \cite{chen2022maat, mehrabi2021survey}. The conceptual gap arises because software fairness is predominantly viewed as purely an ethical issue, with research typically motivated by social requirements, policies, and laws \cite{chen2024fairness, zhang2021ignorance}. This viewpoint overlooks a significant practical benefit: fairness improvement methods can enhance model performance for discriminated groups (unprivileged groups), improve out-of-distribution generalization, and enhance geographical transferability \cite{arjovsky2020out, scholkopf2021toward}. For example, the bias mitigation method FairMask enhances racial fairness in the Correctional Offender Management Profiling for Alternative Sanctions (COMPAS) software system, used by US courts to predict recidivism likelihood \cite{misc_compas_dataset}, by improving predictive performance for Black individuals. Such improvements facilitate the deployment of existing software to predominantly Black regions without needing new systems trained on additional datasets. Recognizing software fairness as inherently tied to software performance helps provide strong motivation for fairness research.

The primary challenge is the trade-off between the applicability and effectiveness of existing bias mitigation methods. Pre-processing methods, which mitigate bias by adjusting training data, are model-agnostic and applicable across ML, and DL. However, recent empirical studies \cite{chen2024fairness, chen2025software} indicate that post-processing methods such as MAAT \cite{chen2022maat}, FairMask \cite{peng2022fairmask}, and MirrorFair \cite{xiao2024mirrorfair} achieve better fairness effectiveness than leading pre-processing methods, such as Fair-SMOTE \cite{chakraborty2021bias} and LTDD \cite{li2022training}, in ML and DL. However, their applicability can be limited by model types due to differences in prediction mechanisms.

To address this dilemma and enable pre-processing methods to achieve state-of-the-art effectiveness in bias mitigation while balancing performance-fairness trade-offs, we propose Correlation Tuning (CoT). CoT adjusts sensitive attribute distributions to tune data correlations explicitly for fairness improvement. Specifically, we utilize the Phi-coefficient \cite{wahono2015systematic, fenton2014software, sultana2017towards, pearson1900x} to measure correlations between sensitive attributes and labels, guiding dataset adjustments to directly mitigate biases associated with sensitive attributes. Additionally, we employ a multi-objective optimization algorithm to address proxy biases caused by non-sensitive attributes \cite{zhang2018fairness, xiao2024mirrorfair}.

We extensively evaluate CoT across ten widely adopted fairness benchmark tasks involving six ML and DL models. The evaluation compares CoT against five state-of-the-art methods across multiple dimensions, including single and multiple sensitive attribute biases, as well as performance-fairness trade-offs and group-level performance. Our results show that CoT improves the true positive rate on unprivileged groups (e.g., ``female'') by an average of 17.5\% across all evaluated tasks without significantly compromising overall performance. For single attributes, CoT reduces SPD, AOD, and EOD biases by 46.9\%, 58.1\%, and 51.0\%, respectively, surpassing existing state-of-the-art methods by more than three percentage points. CoT also outperforms the fairness-performance trade-off baseline in 79.9\% of cases, exceeding state-of-the-art by three percentage points. In addressing multiple sensitive attributes, CoT reduces intersectional biases ISPD, IAOD, and IEOD by 50.3\%, 42.1\%, and 29.5\%, respectively, surpassing existing methods by more than 10 percentage points on average.
\color{black}

Overall, this work makes the following contributions:

\begin{itemize}[leftmargin=10pt]

\item We analyze software fairness, emphasizing that it is both an ethical concern and a \textbf{fundamental software quality issue}. We also highlight the practical significance of fairness research for the SE community, as fairness-improving methods can adjust software performance for specific groups and enhance the out-of-distribution capabilities of software systems.

\item We introduce CoT, a novel pre-processing bias mitigation method that outperforms existing state-of-the-art approaches in fairness improvement and performance-fairness trade-offs.

\item We conduct an extensive empirical evaluation of existing methods, investigating the intersectional impacts of bias mitigation techniques on unconsidered sensitive attributes, which is frequently overlooked in previous fairness research.

\item We will publicly release our data, results, and source code to foster future research in software fairness.

\end{itemize}

\section{Background and Related Work}

In this section, we first introduce the background and key concepts of software fairness, then review the related work.

\subsection{Software Fairness}
\label{sec:bg}

Software fairness concerns fairness bugs in software, which are defined as any imperfections in a software system that cause a discrepancy between actual and required fairness conditions \cite{chen2022fairness, zhang2021ignorance}. Common fairness bugs often arise in classification tasks that assign class labels to individuals based on a set of features \cite{kusner2017counterfactual, biswas2020machine, biswas2023fairify, yu2024fairbalance}. When these tasks are related to significant social activities or important individual benefits, such as college admissions or hiring, the assigned labels can be categorized as either favorable or unfavorable. These are important concepts in software fairness. A \textit{favorable label}, such as ``good credit'', ``high income'', or ``no recidivism'', indicates a potential advantage, for example, increased opportunities to receive benefits, for the recipient. In contrast, an \textit{unfavorable label}, such as ``bad credit'', ``low income'', or ``recidivism'', usually signals potential disadvantages for the individual. Depending on the model's preference for a specific sensitive attribute (also known as protected attribute, e.g., ``sex'', ``age'', and ``race'') when predicting whether an individual receives a favorable or unfavorable label, individuals with a particular sensitive attribute can be categorized into a \textit{privileged group} and an \textit{unprivileged group}. \color{black} From an ethical perspective, model bias reflect human discrimination in the real world. However, from a software quality perspective, it represents a defect, as the model struggles to correctly assign favorable labels to unprivileged groups and unfavorable labels to privileged groups.

% For instance, as mentioned in Section \ref{sec:intro} \cite{misc_compas_dataset}, COMPAS regards White people as the privileged group and tends to assign the favorable label (``no recidivism'') to them, while it regards Black people as the unprivileged group and tends to assign the unfavorable label (``recidivism'') \cite{washington2018argue, propublicaAnalyzedCOMPAS}. 

% From ethical perspective, the model bias continues the human discrimination. However, from software quality perspective, it is defect of hardly assign the correct favorabel label to unprivileged groups and the unfavorabel label to privileged groups.

\color{black}
% Although fairness is well-established ethical issue, software fairness beyond the ethical aspect and is actually a 

\color{black}

% \subsubsection{Software Ethics Perspective} 
% \subsubsection{Software Quality Perspective} 

\subsection{Related Work}
\label{sec:related_work}

With the increasing prevalence of AI-model driven software, concerns regarding fairness have received significant attention in both academia and industry \cite{zhang2021ignorance, zhang2020machine}. As a critical non-functional property tied to software quality and consistency, fairness has garnered heightened interest from the software engineering community \cite{hort2023multi, zhang2022mitigating, li2023faire}. In response, leading technology companies have established dedicated teams to explore equitable AI algorithms, applications, and software services \cite{bird2020fairlearn, madaio2020co, hardt2016equality, beutel2019fairness, sambasivan2021everyone, lahoti2020fairness, prost2019toward}. For instance, IBM has developed the AIF360 toolkit \cite{bellamy2019ai}, which incorporates widely used benchmarking datasets, bias mitigation algorithms, and evaluation metrics.

Bias mitigation strategies are typically categorized into pre-processing, in-processing, and post-processing techniques \cite{mehrabi2021survey}. Specifically, pre-processing methods operate on the data prior to model training by, for example, removing sensitive features, applying instance reweighting, or synthesizing new data samples \cite{zemel2013learning, kamiran2012data}. In contrast, in-processing approaches embed fairness objectives directly into the model learning process, often leveraging techniques such as adversarial debiasing \cite{zhang2018mitigating, beutel2019fairness}. Post-processing techniques, on the other hand, enhance fairness by adjusting the model outputs after training, without modifying the data or the underlying model itself; a representative example is Reject Option Classification \cite{kamishima2012fairness, li2021user}.

While pre-processing methods are widely adopted due to their model-agnostic nature and operational simplicity—particularly in software engineering scenarios \cite{biswas2020machine}—they often face challenges in balancing fairness and overall model performance. To address data bias, Chakraborty et al. \cite{chakraborty2020fairway} propose Fairway, which removes ambiguous data instances to enhance fairness, albeit at the expense of reduced model performance. Subsequently, methods such as Fair-SMOTE \cite{chakraborty2021bias} and FairGenerate \cite{joshi2024fairgenerate} have been introduced to synthesize new data points, aiming for a better trade-off between performance and fairness; however, performance degradation remains a concern. Addressing this limitation, Li et al. \cite{li2022training} present LTDD, which seeks to repair problematic training data rather than simply removing or generating new data.

Regarding in-processing methods, which are applied during the model training phase and are closely linked to model architectures, recent efforts have predominantly focused on deep neural networks (DNNs) \cite{li2023faire}. Sun et al. \cite{sun2022causality} introduce a causality-based approach, CARE, which identifies neurons with high-bias causal effects and subsequently adjusts their activations.

In the context of post-processing, many recent studies have achieved substantial bias mitigation without severely compromising performance. For instance, Peng et al. \cite{peng2022fairmask} propose FairMask, which adjusts sensitive attributes based on an extrapolation model during the model testing stage. Recognizing the dual importance of performance and fairness in AI software, Chen et al. \cite{chen2022maat} present MAAT, which separately trains a performance model and a fairness model, and then ensembles their predictions to achieve fairer outcomes. Similarly, Xiao et al. propose MirrorFair, which generates a counterfactual dataset by flipping sensitive attributes, trains a mirror model with different biases from the original model, and ensembles the predictions of the two models to counteract bias and yield fairer predictions.

Despite these substantial advancements, several challenges remain. Among the three categories of bias mitigation strategies, pre-processing methods are distinguished by their model-agnostic nature. However, they often struggle to achieve an optimal balance between model performance and fairness. In-processing and post-processing methods, while more flexible in overcoming some of the limitations of pre-processing approaches, are typically dependent on specific model architectures or post-processing procedures, which can limit their capacity of generalizing  across different models. For instance, state-of-the-art methods such as MirrorFair are difficult to apply to models that do not explicitly output class probabilities, since they depend on post-prediction probability analysis. To address this limitation, our work proposes a pre-processing method that is model-agnostic and outperforms state-of-the-art approaches in both mitigating bias and balancing model performance with fairness.

\section{Our Approach}
\label{sec:approach}

In this section, we introduce CoT in detail.
% including problem analysis, theoretical derivative and implementation.

\subsection{Overview}
\label{sec:overview}

\begin{figure}[t]
    \centering
    \includegraphics[width=1\linewidth]{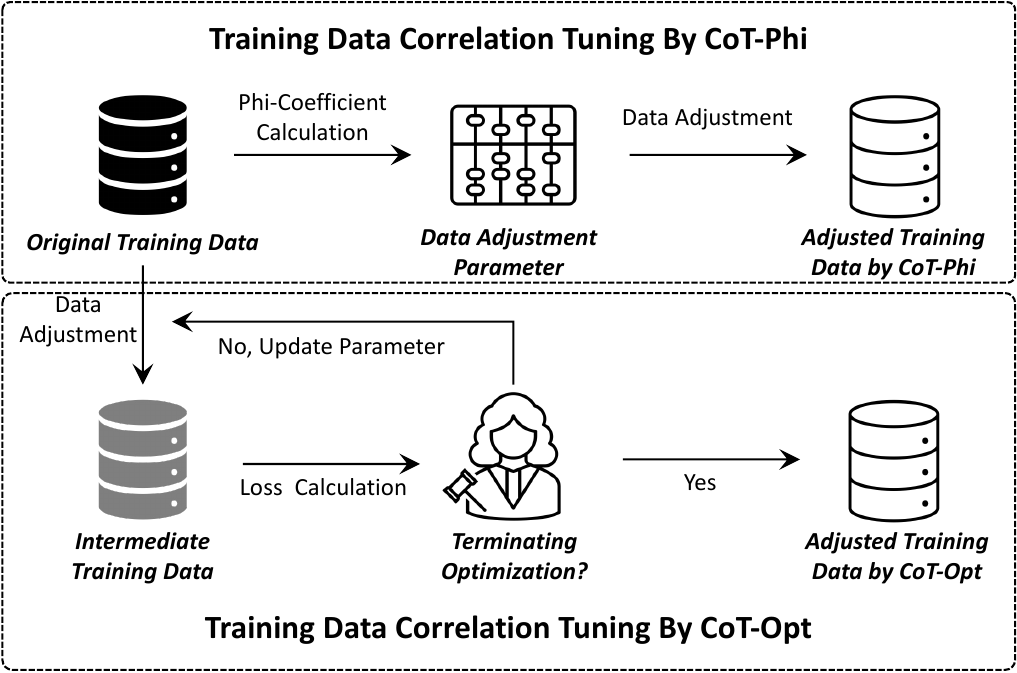}
    \caption{Overview of CoT}
    \label{fig:overview}
\end{figure}

Correlation Tuning (CoT) is a novel pre-processing technique that adjusts the correlation between sensitive attributes and data labels \color{black}to enhance the model performance on unprivileged groups \color{black}to achieve two types of trade-offs: (1) balancing model performance between privileged groups and unprivileged groups, and (2) balancing fairness and performance.  We implement correlation tuning by mutating the values of sensitive attributes. As a classical and intuitive strategy, data features and sensitive attribute adjustment have been applied in other research \cite{feldman2015certifying, calmon2017optimized}. However, the challenge of determining which data instances should be adjusted and how to adjust prevents these methods from significantly improving fairness without excessively compromising performance. Our novelty and main contribution lie in reacting to the ongoing challenge with two separate implementations: (1) CoT via Phi-coefficient tuning, and (2) CoT via multi-objective optimization. Figure \ref{fig:overview} illustrates the overall workflow of the two implementations of CoT, while the details are described in the following sections.

\subsection{Handle Bias by Sensitive Attributes}
\label{sec:sensitive_attribute_bias}
Correlation between data features and labels represents the statistical linear and non-linear relationships between input variables and the target label, which forms the foundation for AI models to make predictions. When a model exhibits bias by disproportionately assigning a favorable label (e.g., ``good credit'') to a privileged group (e.g., ``female''), which is a superficial manifestation. The underlying cause is that the attribute ``female'' displays a stronger statistical association with the label ``good credit'' compared to ``male”, as observed in the training data. Such correlations are determined by the underlying distribution and composition of different groups within the dataset, which may reflect real-world patterns, historical biases, or sampling artifacts.

Therefore, correlation tuning is an approach that addresses bias at its root cause. We choose to adjust the sensitive attribute to tune the correlation rather than oversampling, undersampling, removing specific data instances, synthesizing new data, or modifying data labels. This approach aims to minimize the impact on the original data because the strategies mentioned above alter the correlations among all features and labels, not just between sensitive attributes and data labels. Currently, we lack effective mechanisms to control these negative impacts. More detailed empirical evidence and discussion refer to Section \ref{RQ4} and Section \ref{sec:discussion}. By contrast, tuning correlation through the adjustment of sensitive attributes does not directly affect the correlation between other features and labels. This method gives us greater flexibility in addressing multiple sensitive biases, helping to avoid conflicts and counteracting effects in bias mitigation across different sensitive attributes. In the following section, we describe how we address the two challenges of correlation tuning by adjusting the sensitive attributes.

\subsubsection{Adjust the Sensitive Attribute with Phi-Coefficient}

Consider a prediction task with binary sensitive attributes and labels. Given an imbalanced dataset $D(X, A, Y)$, $X$ represents the non-sensitive attributes, and $A$ is the sensitive attribute taking values $a \in \{0,1\}$, where $a=0$ indicates the unprivileged group and $a=1$ indicates the privileged group. $Y$ denotes the data label, with possible values $y \in \{0,1\}$, where $y=0$ indicates an unfavorable label and $y=1$ indicates a favorable label. Let $I$ represent an instance in $D$, and $n_{A=a,Y=y}$ the number of samples with $A=a$ and $Y=y$.

The goal of correlation tuning is to enhance the positive correlation between the unprivileged group and the favorable label. Existing approaches often globally adjust all data instances to construct a desired class distribution \cite{calmon2017optimized, feldman2015certifying}. For example, Optimized Pre-processing (OP) applies an optimization algorithm to learn a conditional probability transformation matrix that maps the original dataset $D(X, A, Y)$ to a new dataset $D'(X', A', Y')$ for bias mitigation. However, such extensive changes, in the absence of effective risk control mechanisms, hinder the achievement of a promising trade-off between performance and fairness.

Unlike these methods, we select those instances $I_{x, a=1, y=1}$ that belong to the privileged group with a favorable label as candidates for adjustment. The adjustment process simply adjusts selected candidates from privileged to unprivileged as below:

\begin{equation}
    I_{x, a=1, y=1} \longrightarrow I'_{x, a=0, y=1}
\end{equation}

The adjustment preserves the non-sensitive attributes and favorable label, and only changes group membership. This straightforward operation directly alters the class distribution, strengthening the correlation between the unprivileged group and the favorable label. The subsequent challenge is to determine which and how many candidates should be adjusted to achieve a suitable balance between performance and fairness, as well as the performance across both the privileged and unprivileged groups.

\subsubsection{Identify Instances to Be Adjusted by Phi-Coefficient}

To address this problem, we employ the Phi-coefficient \cite{pearson1900x} to determine how many candidates should be adjusted to achieve the desired trade-off. The Phi-coefficient is a significant metric for measuring correlation in binary analysis and is widely used in software engineering literature \cite{wahono2015systematic, fenton2014software, sultana2017towards, pearson1900x}. As a special case of Pearson's coefficient, the Phi-coefficient is particularly well-suited for the context, measuring the correlation between sensitive attributes and data labels. Based on the definitions above, the Phi-coefficient between sensitive attributes and data labels is calculated as follows:

\begin{equation}
\label{eq:Phi-coefficient}
    \phi = \frac{
n_{a=1,\,y=1} \cdot n_{a=0,\,y=0} - n_{a=1,\,y=0} \cdot n_{a=0,\,y=1}
}{
\sqrt{
n_{a=1} \cdot n_{a=0} \cdot n_{y=1} \cdot n_{y=0}
}
}
\end{equation}

Once the Phi-coefficient for the dataset is obtained, its value ranges within $[-1, 1]$. According to our definitions, a positive Phi-coefficient means the privileged group has a stronger correlation with the favorable label (the definition of privileged group aligns with the real situation), while a negative Phi-coefficient means the unprivileged group has a stronger correlation with the favorable label (the definition of privileged group is opposite to the real situation). Ideally, the target Phi-coefficient is \textbf{0}, indicating that both privileged and unprivileged groups have the same correlation to the favorable label. In other words, the sensitive attribute has no correlation with the data labels.

In practice, the Phi-coefficient of a real dataset is usually not zero. Instead, based on our definitions, it is typically positive. Therefore, the goal of correlation tuning is to adjust the Phi-coefficient to zero by transforming candidate instances from $I_{x, a=1, y=1}$ to $I'_{x, a=0, y=1}$. Let $P_{\phi} \in (0,1)$ be the adjustment parameter for the Phi-coefficient and $k$ denote the number of adjusted instances. Using Eq.~(\ref{eq:Phi-coefficient}), we can derive the analytical solution for $P_{\phi}$ as follows:

\begin{equation}
\label{eq:P_Phi}
    P_{\phi} = \frac{k}{n_{a=1, y=1}} 
    = \frac{n_{a=1, y=1} \cdot n_{a=0, y=0} - n_{a=1, y=0} \cdot n_{a=0, y=1}}{n_{a=1, y=1} \cdot n_{a=0, y=0} n_{a=1, y=0}}
\end{equation}

With $P_{\phi}$, we randomly select the corresponding proportion of candidates for instance adjustment, transforming the dataset from $D(X, A, Y)$ to $D'(X, A', Y)$ and completing the correlation tuning process based on the Phi-coefficient. We refer to this implementation of CoT as \textbf{CoT-Phi}.

\subsection{Handle Bias by Non-sensitive Attributes}
\label{sec:non-sensitive_attribute_bias}

Through correlation tuning via CoT-Phi, we explicitly adjust the correlation between sensitive attributes and labels to a desired value. However, many previous works have revealed that model bias can arise not only from the sensitive attributes themselves, but also from the proxy effects of non-sensitive attributes \cite{xiao2024mirrorfair, zhang2018fairness, pearl2018book, pearl2009causality, datta2017proxy, googleMachineLearning-proxy}. Currently, we lack effective and convenient tools to analyze the correlation among multiple variables, especially when these variables include both categorical and numerical types. 

To address the influence caused by non-sensitive attributes, we propose a compensating strategy that dynamically adjusts the Phi-coefficient between sensitive attributes and labels, rather than setting it to zero by default. This design allows us to counteract the effect of non-sensitive attributes by strengthening or weakening the correlation between sensitive attributes and labels. For example, we may build a stronger correlation between the unprivileged group and the favorable label to offset the bias introduced by non-sensitive attributes.

To achieve this, we introduce multi-objective optimization \cite{yu2022towards, censor1977pareto, chaudhari2022simultaneous} into our approach to identify the optimal Phi-coefficient value that achieves a balance between model performance and fairness, as well as balanced performance across groups. We define the loss function for the multi-objective optimization process as follows:

\vspace{-1.5em}
\begin{align}
\label{eq:loss_function}
f(P_\phi) &= [1-f_1(P_\phi)] + [1-f_2(P_\phi)] + f_3(P_\phi) + f_4(P_\phi)\\ \nonumber & + f_5(P_\phi), \quad P_\phi \in (0, 1)
\end{align}

Here, $P_\phi$ denotes the proportion of candidate instances to be adjusted. $f_{1}$ denotes the F1 Score metric, $f_{2}$ denotes the accuracy metric, $f_{3}$ denotes the SPD metric, $f_{4}$ denotes the AOD metric, and $f_{5}$ denotes the EOD metric. This loss function design comprehensively incorporates both performance and fairness metrics to achieve a better trade-off. Specifically, F1 score and accuracy measure model performance, while SPD, AOD, and EOD serve as metrics of model fairness. Details of these metrics are provided in Section \ref{sec:metrics_measurements}, and further discussion on the loss function design can be found in Section \ref{sec:discussion}. 

We refer to this implementation as \textbf{CoT-Opt}. \color{black} CoT-Opt can be regarded as a hybrid that combines the simplicity of CoT-Phi with the flexibility of optimization. The key distinction between CoT-Phi and CoT-Opt lies in how the tuning parameter is determined: CoT-Phi fixes it analytically, whereas CoT-Opt dynamically optimizes it with respect to different loss functions. Thus, CoT-Opt is designed as an optimized extension of CoT-Phi. \color{black}

\subsection{Handle Multiple Sensitive Attributes Bias}

In real-world AI decision-making scenarios, it is common to encounter cases where more than one sensitive attribute needs bias mitigation. For example, in the COMPAS software application \cite{propublicaAnalyzedCOMPAS}, both ``sex'' and ``race'' are class-sensitive attributes that require protection and mitigation of related biases. As a method focused on adjusting sensitive attributes, CoT offers specific advantages in addressing multiple sensitive attributes.

\subsubsection{CoT-Phi for Multiple Sensitive Attributes}

We can repeat the process used for a single attribute to mitigate multiple sensitive attribute biases. Given an original dataset $D(X, A1, A2, Y)$, where $A1$ represents ``sensitive attribute 1'' and $A2$ represents ``sensitive attribute 2,'' the adjustment process can be described as follows, using parameters $P_{\phi1}$ and $P_{\phi2}$ derived from Eq (\ref{eq:P_Phi}):

\begin{equation}
    D(X, A1, A2, Y) \rightarrow D'(X, A1', A2, Y) \rightarrow D''(X, A1', A2', Y)
\end{equation}

\subsubsection{CoT-Opt for Multiple Sensitive Attributes}

Similarly, we apply the same approach as with a single sensitive attribute using CoT-Opt to adjust the dataset from $D(X, A1, A2, Y)$ to $D'(X, A1', A2, Y)$. We then introduce the following loss function:

\begin{align}
\label{eq:loss_function_2}
f(P_\phi) &= [1-f_1(P_\phi)] + [1-f_2(P_\phi)] + f_6(P_\phi) + f_7(P_\phi)\\ \nonumber & + f_8(P_\phi), \quad P_\phi \in (0, 1)
\end{align}

Compared to the previous loss function (\ref{eq:loss_function}), we replace the single attribute fairness metrics (SPD, AOD, and EOD) with intersectional fairness metrics (ISPD, IAOD, and IEOD; for detailed introductions to these metrics, refer to Section \ref{tab:fairness_metrics}). This design enables a better balance between mitigating different types of bias and maintaining performance. Additional discussion on this design is provided in Section \ref{sec:discussion}. With this, we have introduced the methodology of CoT, and next, we move to the evaluation section.

\section{Evaluation}
\label{sec:eval}
% \revise{In this section, we introduce the experiment design, metrics, and come up with our research questions.}\xiancm{you have delete the research questions, pls rewrite this part, I have give an example here.}
In this part, we introduce our experimental design to evaluate the performance and applicability of CoT by answering the following research questions (RQs). We present the benchmark datasets and evaluation metrics and then provide detailed experiment setups. 

\subsection{Research Questions}
% \subsubsection{RQ1(Effectiveness):}
% \subsubsubsection 

% The effectiveness of FITNESS was evaluated via answering the following research questions.
\begin{itemize}[leftmargin = 10 pt]

\color{black}
\item \textbf{RQ1:} \textit{What is the impact of CoT on model performance and fairness?}
This RQ investigates the quantitative impact of CoT on model performance and fairness by comparing various metrics across models applied to two implementations of CoT and the original model without applying any bias mitigation methods. \color{black}
   
\item \textbf{RQ2:} \textit{What is the bias mitigation effectiveness of CoT compared with existing methods?} 
This research question examines the advantages of CoT in bias mitigation compared to existing methods by analyzing both the absolute values and relative changes of SPD, AOD, and EOD fairness metrics across various tasks.

\item \textbf{RQ3:} \textit{What is the performance-fairness trade-off of CoT compared with existing methods?}
This RQ investigates the advantages of CoT in balancing model performance and fairness compared to existing methods, using the Fairea Trade-off measurement tool.

\item \textbf{RQ4:} \textit{What is the effectiveness of CoT in handling multiple sensitive attributes?}
This RQ examines the potential negative side effects of fairness-improving methods on unconsidered sensitive attributes and evaluates the effectiveness of CoT in mitigating multiple sensitive attribute biases and intersectional bias.

\color{black}
\item \textbf{RQ5:} \textit{What is the robustness of CoT under overfitting risks and realistic data conditions?}  
This RQ investigates the robustness of CoT by evaluating whether its effectiveness remain consistent when facing overfitting, missing values, noise, and outliers.
\color{black}
% \item \textbf{RQ5:} 

% How do multi-objective optimization strategies influence the efficacy of FITNESS? This research question involves assigning varying weights to distinct performance and fairness metrics within the multi-objective optimization loss function to delve into the impact of diverse multi-objective optimization approaches on FITNESS's effectiveness.
    
\end{itemize}
% MacBook Pro with 16GB RAM, M1 Pro CPU/GPU.
%XXX\revise{fill you devices details, e.g., CPU @3.00GHZ and 215G memory ....} 

% \begin{figure}[!t]
% \centering 
% \vspace{-1ex}
% \includegraphics[width=\linewidth]{fig/distribution.pdf}
% \vspace{-2ex}
% \caption{The distribution of experiment benchmark datasets}
% %\xiancm{pls change the caption, Maybe the experiment dataset distribution would be better}
% \vspace{-2ex}
% \label{fig:sample distribution} 
% \end{figure} 

%\subsection{Benchmark Datasets and Terminologies}

\begin{table}[h]
  \centering
  \caption{Benchmark datasets.}
  \vspace{-2ex}
  \resizebox{\columnwidth}{!}{
% Table generated by Excel2LaTeX from sheet 'Datasets_Info'
    \begin{tabular}{lrlrrr}
    \hline
    Dataset & Feature Size & Sensitive Attributes & Data Size & Favorable Label & Mojority Label \bigstrut\\
    \hline
    Adult  & 7    & sex, race & 48,842 & 1 (income > 50K ) & 0 (76.1\%) \bigstrut[t]\\
    Compas & 7    & sex, race & 7,214 & 0 (no recidivism) & 0 (54.9\%) \\
    Default & 23    & sex, age & 30,000 & 0 (no default) & 0 (77.9\%) \\
    Mep1  & 42    & sex, race   & 15,830 & 1 (subscriber) & 0 (82.8\%) \\
    MEP2   & 42 & sex, race  & 15,675 & 1 (utilizer) & 0 (82.2\%) \bigstrut[b]\\
    \hline
    \end{tabular}%
    }
  \label{table 1}%
  \vspace{-1em}
\end{table}%

\subsection{Benchmark Datasets}
\label{sec:experiment:dataset}
Following recent empirical studies on the fairness of machine learning software, we conducted our experiments on five benchmark datasets using the IBM AIF360 framework \cite{chen2024fairness}. These datasets originate from diverse domains, including finance, criminal justice, and healthcare, and have been widely adopted within the fairness research community.

The Adult Income dataset \cite{misc_adult_2} (also known as the Adult dataset) contains individual information derived from the 1994 U.S. Census, with the objective of predicting whether an individual's income exceeds \$50K. The ProPublica Recidivism dataset \cite{misc_compas_dataset} (also referred to as the Compas dataset) comprises the criminal history of defendants, aiming to predict their potential for re-offending in the future. The Default of Credit Card Clients dataset \cite{uciMachineLearning-default} (also referred to as the Default dataset) encompasses information on default payments, demographic factors, credit data, payment history, and bill statements of credit card clients. Lastly, the Medical Survey 2015 and Medical Survey 2016 datasets \cite{ahrqMedicalExpenditure-mep, ahrqMedicalExpenditure-mep2} (also known as the Mep1 and Mep2 datasets) contain data regarding American medical care usage, health insurance, and out-of-pocket expenditures, with the goal of predicting healthcare utilization patterns.

\color{black}
These benchmark datasets are well-established and do not require additional pre-processing or cleaning for fairness research. They are particularly valuable as they comprehensively represent real-world scenarios and have been adopted in numerous significant studies \cite{chen2024fairness, xiao2024mirrorfair, chen2025software, chen2024diversity}, providing meaningful insights into the equity and ethical implications of machine learning algorithms.
\color{black}

\subsection{Existing Methods}
We adopt five state-of-the-art bias mitigation methods as baselines to demonstrate the effectiveness of CoT, including three pre-processing and two non-preprocessing techniques. These existing methods are derived from recent, significant fairness research published in flagship venues. Fair-SMOTE (FSE'21) \cite{chakraborty2021bias} generates synthetic data using the SMOTE \cite{fernandez2018smote, chawla2002smote} algorithm to mitigate imbalances in the training distribution and reduce bias. LTDD (ICSE'22) \cite{li2022training} enhances fairness by learning fair data representations through a transfer learning framework. FairMask (TSE'23) \cite{peng2022fairmask} addresses bias by substituting sensitive attributes in the test set with synthesized values. MirrorFair (FSE'24) \cite{xiao2024mirrorfair} creates counterfactual samples by flipping sensitive attributes, trains a mirror model, and ensembles predictions from both models for fairer outcomes. FairGenerate (TOSEM'25) \cite{joshi2024fairgenerate} leverages generative modeling to augment training data and improve fairness.

\subsection{Evaluation Metrics and Measurements}
\label{sec:metrics_measurements}
\subsubsection{Performance Metrics}
We use five widely adopted performance metrics: precision, recall, accuracy, F1-score, and Matthews correlation coefficient (MCC). Table \ref{tab:performance_matrics} presents the definitions of the model performance metrics used in this paper. The notation $TP$ signifies \textit{True Positives}, $TN$ signifies \textit{True Negatives}, $FP$ signifies \textit{False Positives}, and $FN$ signifies \textit{False Negatives}. Precision measures the exactness of the model in predicting a certain class; recall measures the sensitivity of the model to a certain class; accuracy measures the overall correctness of model predictions; F1-score is the harmonic mean of precision and recall. MCC measures the correlation between observed and predicted classifications, providing a balanced evaluation even when classes are imbalanced.

% Table generated by Excel2LaTeX from sheet 'baselines'
\begin{table}[htbp]
  \centering
  % \renewcommand{\arraystretch}{1.8}
  % \footnotesize 
  \caption{Definition of model performance metrics. }
  \resizebox{\linewidth}{!}{
    \begin{tabular}{l|c}
    \toprule
    Metrics & Definition \bigstrut[t]\\
    \midrule
    Precision   &  ${TP}/(TP + FP)$ \\
    % \hline
    Recall   &  $ {TP}/(TP + FN)$\\
    % \hline
    Accuracy   & $(TP + TN)/(TP+FP+TN+FN)$ \\
    % \hline
    F1-Score & ${2 \times Precision \times Recall}/(Precision + Recall)$ \\
    % \hline
    MCC     &  $ (TP \times TN - FP \times FN)/{\sqrt{(TP+FP)(TP+FN)(TN+FP)(TN+FN)}}$ \\
    \bottomrule
    \end{tabular}%
    }
    \vspace{-1em}
  \label{tab:performance_matrics}%
\end{table}%

\subsubsection{Fairness Metrics}
To comprehensively evaluate both single and intersectional fairness, we adopt three widely used fairness metrics: Statistical Parity Difference, Average Odds Difference, and Equal Opportunity Difference, along with their intersectional counterparts (ISPD, IAOD, IEOD), following prior works~\cite{chen2022maat, chen2023comprehensive, biswas2020machine, chen2024fairness}. Specifically, SPD measures the difference in favorable prediction rates between unprivileged and privileged groups; AOD is the average of the differences in False Positive Rates (FPR) and True Positive Rates (TPR) across groups; EOD quantifies the TPR gap between unprivileged and privileged groups. \color{black} The measurement of these fairness metrics actually evaluates the performance differences across groups, further reflecting the nature of software fairness issues—namely, a quality defect characterized by performance disparities among user groups.\color{black}

The formal definitions of these metrics are presented in Table~\ref{tab:fairness_metrics}.
Where $A$ denotes the sensitive attribute, $Y$ the ground-truth label, and $\hat{Y}$ the predicted label. $A=0$ and $A=1$ indicate unprivileged and privileged groups, respectively, while $Y=1$ represents the favorable outcome. For intersectional metrics, $s \in S$ enumerates all intersectional subgroups (e.g., gender and race combinations).  For intersectional fairness, the metrics (ISPD, IAOD, IEOD) extend these notions to the worst-case disparity among all intersectional subgroups $S$, capturing fairness at the granularity of sensitive attribute intersections. For instance, in a decision-making task where ``sex'' and ``race'' are sensitive attributes to be protected, if ``male'' and ``White'' are privileged groups while ``female'' and ``Non-White'' are unprivileged groups, intersectional fairness measures the disparities between the ``male-White'' and ``female-Non-White'' subgroups.

% Table generated by Excel2LaTeX from sheet 'baselines'
\begin{table}[htbp]
  \centering
  \renewcommand{\arraystretch}{1.8}
  % \footnotesize 
  \caption{Definition of single and intersectional fairness bias. }
  \resizebox{\linewidth}{!}{
    \begin{tabular}{l|c}
    \toprule
    Metrics & Definition \bigstrut[t]\\
    \midrule
    SPD   &  $P[\hat{Y} = 1\,|\,A = 0] - P[\hat{Y} = 1\,|\,A = 1]$ \\
    % \hline
    AOD   &  $\begin{aligned}
        &\frac{1}{2}(P[\hat{Y} = 1 \mid A = 0, Y = 0] - P[\hat{Y} = 1 \mid A = 1, Y = 0] \\
                   &+\, P[\hat{Y} = 1 \mid A = 0, Y = 1] - P[\hat{Y} = 1 \mid A = 1, Y = 1])
        \end{aligned}$\\
    % \hline
    EOD   & $P[\hat{Y} = 1\,|\,A = 0, Y = 1] - P[\hat{Y} = 1\,|\,A = 1, Y = 1]$ \bigstrut[b]\\
    \hline
    ISPD & $\max_{s \in S} P[\hat{Y} = 1 \mid A = s] - \min_{s \in S} P[\hat{Y} = 1 \mid A = s]$ \\
    % \hline
    IAOD &
    \small
    $\begin{aligned}
    &\frac{1}{2}[\max_{s \in S}(P[\hat{Y} = 1 \mid A = s, Y = 0] + P[\hat{Y} = 1 \mid A = s, Y = 1]) \\
        &- \min_{s \in S}(P[\hat{Y} = 1 \mid A = s, Y = 0] + P[\hat{Y} = 1 \mid A = s, Y = 1])]
    \end{aligned}$
    \\
    % \hline
    IEOD & $\begin{aligned}\max_{s \in S} P[\hat{Y} = 1 \mid A = s, Y = 1] - \min_{s \in S} P[\hat{Y} = 1 \mid A = s, Y = 1]\end{aligned}$ \\
    \bottomrule

    \end{tabular}%
    }
  \label{tab:fairness_metrics}%
\end{table}%

% Table generated by Excel2LaTeX from sheet 'Sheet4'
\begin{table*}[!ht]
  \centering
  \caption{(RQ1) Impact of CoT on model performance and fairness. TPR-U denotes the true positive rate of the unprivileged group and Acc denotes model accuracy. ``+'' indicates that higher values are better, while ``–'' indicates that lower values are better. \color{black} We use \colorbox[rgb]{ .761,  .831,  .839}{Teal Background} to highlight the TPR-U, and \colorbox[rgb]{ .902,  .671,  .608}{Salmon Background} to highlight the SPD, AOD, and EOD after applying CoT-Phi and CoT-Opt. CoT-Phi and CoT-Opt increase TPR-U by 17\% and 15\%, respectively, and reduce SPD, AOD, and EOD by 57\%/46\%, 55\%/58\%, and 41\%/51\%, respectively. \color{black} }

  \resizebox{\linewidth}{!}{
\begin{tabular}{l|l|rrrrr|rrrrr|rrrrr|rrrrr|rrrrr|rrrrr}
\hline
\multirow{4}[6]{*}{\textbf{Task}} & \multirow{4}[6]{*}{\textbf{Method}} & \multicolumn{5}{c|}{\cellcolor[rgb]{ .651,  .651,  .651}\textbf{LR}} & \multicolumn{5}{c|}{\cellcolor[rgb]{ .651,  .651,  .651}\textbf{SVM}} & \multicolumn{5}{c|}{\cellcolor[rgb]{ .651,  .651,  .651}\textbf{XGB}} & \multicolumn{5}{c|}{\cellcolor[rgb]{ .651,  .651,  .651}\textbf{LGBM}} & \multicolumn{5}{c|}{\cellcolor[rgb]{ .651,  .651,  .651}\textbf{RF}} & \multicolumn{5}{c}{\cellcolor[rgb]{ .651,  .651,  .651}\textbf{DNN}} \bigstrut\\
\cline{3-32}      &       & \multicolumn{2}{c}{\cellcolor[rgb]{ .651,  .651,  .651}\textbf{Performance}} & \multicolumn{3}{c|}{\cellcolor[rgb]{ .651,  .651,  .651}\textbf{Fairness}} & \multicolumn{2}{c}{\cellcolor[rgb]{ .651,  .651,  .651}\textbf{Performance}} & \multicolumn{3}{c|}{\cellcolor[rgb]{ .651,  .651,  .651}\textbf{Fairness}} & \multicolumn{2}{c}{\cellcolor[rgb]{ .651,  .651,  .651}\textbf{Performance}} & \multicolumn{3}{c|}{\cellcolor[rgb]{ .651,  .651,  .651}\textbf{Fairness}} & \multicolumn{2}{c}{\cellcolor[rgb]{ .651,  .651,  .651}\textbf{Performance}} & \multicolumn{3}{c|}{\cellcolor[rgb]{ .651,  .651,  .651}\textbf{Fairness}} & \multicolumn{2}{c}{\cellcolor[rgb]{ .651,  .651,  .651}\textbf{Performance}} & \multicolumn{3}{c|}{\cellcolor[rgb]{ .651,  .651,  .651}\textbf{Fairness}} & \multicolumn{2}{c}{\cellcolor[rgb]{ .651,  .651,  .651}\textbf{Performance}} & \multicolumn{3}{c}{\cellcolor[rgb]{ .651,  .651,  .651}\textbf{Fairness}} \bigstrut[t]\\
      &       & \multicolumn{2}{c}{\cellcolor[rgb]{ .651,  .651,  .651}\textbf{ (+)}} & \multicolumn{3}{c|}{\cellcolor[rgb]{ .651,  .651,  .651}\textbf{ (-)}} & \multicolumn{2}{c}{\cellcolor[rgb]{ .651,  .651,  .651}\textbf{ (+)}} & \multicolumn{3}{c|}{\cellcolor[rgb]{ .651,  .651,  .651}\textbf{ (-)}} & \multicolumn{2}{c}{\cellcolor[rgb]{ .651,  .651,  .651}\textbf{ (+)}} & \multicolumn{3}{c|}{\cellcolor[rgb]{ .651,  .651,  .651}\textbf{ (-)}} & \multicolumn{2}{c}{\cellcolor[rgb]{ .651,  .651,  .651}\textbf{ (+)}} & \multicolumn{3}{c|}{\cellcolor[rgb]{ .651,  .651,  .651}\textbf{ (-)}} & \multicolumn{2}{c}{\cellcolor[rgb]{ .651,  .651,  .651}\textbf{ (+)}} & \multicolumn{3}{c|}{\cellcolor[rgb]{ .651,  .651,  .651}\textbf{ (-)}} & \multicolumn{2}{c}{\cellcolor[rgb]{ .651,  .651,  .651}\textbf{ (+)}} & \multicolumn{3}{c}{\cellcolor[rgb]{ .651,  .651,  .651}\textbf{ (-)}} \bigstrut[b]\\
\cline{3-32}      &       & \cellcolor[rgb]{ .651,  .651,  .651}\textbf{TPR-U} & \cellcolor[rgb]{ .651,  .651,  .651}\textbf{Acc} & \cellcolor[rgb]{ .651,  .651,  .651}\textbf{SPD} & \cellcolor[rgb]{ .651,  .651,  .651}\textbf{AOD} & \cellcolor[rgb]{ .651,  .651,  .651}\textbf{EOD} & \cellcolor[rgb]{ .651,  .651,  .651}\textbf{TPR-U} & \cellcolor[rgb]{ .651,  .651,  .651}\textbf{Acc} & \cellcolor[rgb]{ .651,  .651,  .651}\textbf{SPD} & \cellcolor[rgb]{ .651,  .651,  .651}\textbf{AOD} & \cellcolor[rgb]{ .651,  .651,  .651}\textbf{EOD} & \cellcolor[rgb]{ .651,  .651,  .651}\textbf{TPR-U} & \cellcolor[rgb]{ .651,  .651,  .651}\textbf{Acc} & \cellcolor[rgb]{ .651,  .651,  .651}\textbf{SPD} & \cellcolor[rgb]{ .651,  .651,  .651}\textbf{AOD} & \cellcolor[rgb]{ .651,  .651,  .651}\textbf{EOD} & \cellcolor[rgb]{ .651,  .651,  .651}\textbf{TPR-U} & \cellcolor[rgb]{ .651,  .651,  .651}\textbf{Acc} & \cellcolor[rgb]{ .651,  .651,  .651}\textbf{SPD} & \cellcolor[rgb]{ .651,  .651,  .651}\textbf{AOD} & \cellcolor[rgb]{ .651,  .651,  .651}\textbf{EOD} & \cellcolor[rgb]{ .651,  .651,  .651}\textbf{TPR-U} & \cellcolor[rgb]{ .651,  .651,  .651}\textbf{Acc} & \cellcolor[rgb]{ .651,  .651,  .651}\textbf{SPD} & \cellcolor[rgb]{ .651,  .651,  .651}\textbf{AOD} & \cellcolor[rgb]{ .651,  .651,  .651}\textbf{EOD} & \cellcolor[rgb]{ .651,  .651,  .651}\textbf{TPR-U} & \cellcolor[rgb]{ .651,  .651,  .651}\textbf{Acc} & \cellcolor[rgb]{ .651,  .651,  .651}\textbf{SPD} & \cellcolor[rgb]{ .651,  .651,  .651}\textbf{AOD} & \cellcolor[rgb]{ .651,  .651,  .651}\textbf{EOD} \bigstrut\\
\hline
\multirow{3}[2]{*}{\textbf{\makecell[l]{Adult\\Sex}}} & \textbf{Original} & \textbf{0.18} & \textbf{0.82} & \textbf{0.17} & \textbf{0.17} & \textbf{0.27} & \textbf{0.22} & \textbf{0.82} & \textbf{0.13} & \textbf{0.12} & \textbf{0.19} & \textbf{0.34} & \textbf{0.85} & \textbf{0.18} & \textbf{0.15} & \textbf{0.23} & \textbf{0.33} & \textbf{0.85} & \textbf{0.18} & \textbf{0.15} & \textbf{0.24} & \textbf{0.33} & \textbf{0.84} & \textbf{0.18} & \textbf{0.15} & \textbf{0.22} & \textbf{0.23} & \textbf{0.83} & \textbf{0.19} & \textbf{0.18} & \textbf{0.28} \bigstrut[t]\\
      & \textbf{CoT-Phi} & \cellcolor[rgb]{ .761,  .831,  .839}\textbf{0.36} & \textbf{0.81} & \cellcolor[rgb]{ .902,  .671,  .608}\textbf{0.06} & \cellcolor[rgb]{ .902,  .671,  .608}\textbf{0.01} & \cellcolor[rgb]{ .902,  .671,  .608}\textbf{0.02} & \cellcolor[rgb]{ .761,  .831,  .839}\textbf{0.35} & \textbf{0.81} & \cellcolor[rgb]{ .902,  .671,  .608}\textbf{0.06} & \cellcolor[rgb]{ .902,  .671,  .608}\textbf{0.01} & \cellcolor[rgb]{ .902,  .671,  .608}\textbf{0.02} & \cellcolor[rgb]{ .761,  .831,  .839}\textbf{0.47} & \textbf{0.85} & \cellcolor[rgb]{ .902,  .671,  .608}\textbf{0.09} & \cellcolor[rgb]{ .902,  .671,  .608}\textbf{0.01} & \cellcolor[rgb]{ .902,  .671,  .608}\textbf{0.02} & \cellcolor[rgb]{ .761,  .831,  .839}\textbf{0.48} & \textbf{0.84} & \cellcolor[rgb]{ .902,  .671,  .608}\textbf{0.09} & \cellcolor[rgb]{ .902,  .671,  .608}\textbf{0.01} & \cellcolor[rgb]{ .902,  .671,  .608}\textbf{0.01} & \cellcolor[rgb]{ .761,  .831,  .839}\textbf{0.47} & \textbf{0.83} & \cellcolor[rgb]{ .902,  .671,  .608}\textbf{0.07} & \cellcolor[rgb]{ .902,  .671,  .608}\textbf{0.01} & \cellcolor[rgb]{ .902,  .671,  .608}\textbf{0.02} & \cellcolor[rgb]{ .761,  .831,  .839}\textbf{0.41} & \textbf{0.82} & \cellcolor[rgb]{ .902,  .671,  .608}\textbf{0.07} & \cellcolor[rgb]{ .902,  .671,  .608}\textbf{0.03} & \cellcolor[rgb]{ .902,  .671,  .608}\textbf{0.04} \\
      & \textbf{CoT-Opt} & \cellcolor[rgb]{ .761,  .831,  .839}\textbf{0.37} & \textbf{0.81} & \cellcolor[rgb]{ .902,  .671,  .608}\textbf{0.06} & \cellcolor[rgb]{ .902,  .671,  .608}\textbf{0.01} & \cellcolor[rgb]{ .902,  .671,  .608}\textbf{0.02} & \cellcolor[rgb]{ .761,  .831,  .839}\textbf{0.36} & \textbf{0.81} & \cellcolor[rgb]{ .902,  .671,  .608}\textbf{0.06} & \cellcolor[rgb]{ .902,  .671,  .608}\textbf{0.01} & \cellcolor[rgb]{ .902,  .671,  .608}\textbf{0.03} & \cellcolor[rgb]{ .761,  .831,  .839}\textbf{0.48} & \textbf{0.84} & \cellcolor[rgb]{ .902,  .671,  .608}\textbf{0.08} & \cellcolor[rgb]{ .902,  .671,  .608}\textbf{0.01} & \cellcolor[rgb]{ .902,  .671,  .608}\textbf{0.02} & \cellcolor[rgb]{ .761,  .831,  .839}\textbf{0.48} & \textbf{0.84} & \cellcolor[rgb]{ .902,  .671,  .608}\textbf{0.08} & \cellcolor[rgb]{ .902,  .671,  .608}\textbf{0.01} & \cellcolor[rgb]{ .902,  .671,  .608}\textbf{0.02} & \cellcolor[rgb]{ .761,  .831,  .839}\textbf{0.47} & \textbf{0.83} & \cellcolor[rgb]{ .902,  .671,  .608}\textbf{0.07} & \cellcolor[rgb]{ .902,  .671,  .608}\textbf{0.01} & \cellcolor[rgb]{ .902,  .671,  .608}\textbf{0.02} & \cellcolor[rgb]{ .761,  .831,  .839}\textbf{0.42} & \textbf{0.82} & \cellcolor[rgb]{ .902,  .671,  .608}\textbf{0.07} & \cellcolor[rgb]{ .902,  .671,  .608}\textbf{0.03} & \cellcolor[rgb]{ .902,  .671,  .608}\textbf{0.04} \bigstrut[b]\\
\hline
\multirow{3}[2]{*}{\textbf{\makecell[l]{Adult\\Race}}} & \textbf{Original} & \textbf{0.30} & \textbf{0.82} & \textbf{0.09} & \textbf{0.08} & \textbf{0.12} & \textbf{0.32} & \textbf{0.82} & \textbf{0.07} & \textbf{0.05} & \textbf{0.07} & \textbf{0.45} & \textbf{0.85} & \textbf{0.09} & \textbf{0.06} & \textbf{0.09} & \textbf{0.44} & \textbf{0.85} & \textbf{0.09} & \textbf{0.07} & \textbf{0.10} & \textbf{0.45} & \textbf{0.84} & \textbf{0.09} & \textbf{0.06} & \textbf{0.08} & \textbf{0.37} & \textbf{0.82} & \textbf{0.09} & \textbf{0.06} & \textbf{0.09} \bigstrut[t]\\
      & \textbf{CoT-Phi} & \cellcolor[rgb]{ .761,  .831,  .839}\textbf{0.45} & \textbf{0.82} & \cellcolor[rgb]{ .902,  .671,  .608}\textbf{0.04} & \cellcolor[rgb]{ .902,  .671,  .608}\textbf{0.02} & \cellcolor[rgb]{ .902,  .671,  .608}\textbf{0.05} & \cellcolor[rgb]{ .761,  .831,  .839}\textbf{0.41} & \textbf{0.82} & \cellcolor[rgb]{ .902,  .671,  .608}\textbf{0.03} & \cellcolor[rgb]{ .902,  .671,  .608}\textbf{0.02} & \cellcolor[rgb]{ .902,  .671,  .608}\textbf{0.04} & \cellcolor[rgb]{ .761,  .831,  .839}\textbf{0.56} & \textbf{0.85} & \cellcolor[rgb]{ .902,  .671,  .608}\textbf{0.05} & \cellcolor[rgb]{ .902,  .671,  .608}\textbf{0.02} & \cellcolor[rgb]{ .902,  .671,  .608}\textbf{0.03} & \cellcolor[rgb]{ .761,  .831,  .839}\textbf{0.55} & \textbf{0.85} & \cellcolor[rgb]{ .902,  .671,  .608}\textbf{0.05} & \cellcolor[rgb]{ .902,  .671,  .608}\textbf{0.02} & \cellcolor[rgb]{ .902,  .671,  .608}\textbf{0.03} & \cellcolor[rgb]{ .761,  .831,  .839}\textbf{0.55} & \textbf{0.84} & \cellcolor[rgb]{ .902,  .671,  .608}\textbf{0.04} & \cellcolor[rgb]{ .902,  .671,  .608}\textbf{0.02} & \cellcolor[rgb]{ .902,  .671,  .608}\textbf{0.04} & \cellcolor[rgb]{ .761,  .831,  .839}\textbf{0.50} & \textbf{0.83} & \cellcolor[rgb]{ .902,  .671,  .608}\textbf{0.03} & \cellcolor[rgb]{ .902,  .671,  .608}\textbf{0.04} & \cellcolor[rgb]{ .902,  .671,  .608}\textbf{0.06} \\
      & \textbf{CoT-Opt} & \cellcolor[rgb]{ .761,  .831,  .839}\textbf{0.42} & \textbf{0.82} & \cellcolor[rgb]{ .902,  .671,  .608}\textbf{0.05} & \cellcolor[rgb]{ .902,  .671,  .608}\textbf{0.02} & \cellcolor[rgb]{ .902,  .671,  .608}\textbf{0.03} & \cellcolor[rgb]{ .761,  .831,  .839}\textbf{0.40} & \textbf{0.82} & \cellcolor[rgb]{ .902,  .671,  .608}\textbf{0.04} & \cellcolor[rgb]{ .902,  .671,  .608}\textbf{0.01} & \cellcolor[rgb]{ .902,  .671,  .608}\textbf{0.03} & \cellcolor[rgb]{ .761,  .831,  .839}\textbf{0.56} & \textbf{0.85} & \cellcolor[rgb]{ .902,  .671,  .608}\textbf{0.05} & \cellcolor[rgb]{ .902,  .671,  .608}\textbf{0.02} & \cellcolor[rgb]{ .902,  .671,  .608}\textbf{0.04} & \cellcolor[rgb]{ .761,  .831,  .839}\textbf{0.56} & \textbf{0.85} & \cellcolor[rgb]{ .902,  .671,  .608}\textbf{0.05} & \cellcolor[rgb]{ .902,  .671,  .608}\textbf{0.02} & \cellcolor[rgb]{ .902,  .671,  .608}\textbf{0.04} & \cellcolor[rgb]{ .761,  .831,  .839}\textbf{0.54} & \textbf{0.84} & \cellcolor[rgb]{ .902,  .671,  .608}\textbf{0.05} & \cellcolor[rgb]{ .902,  .671,  .608}\textbf{0.02} & \cellcolor[rgb]{ .902,  .671,  .608}\textbf{0.03} & \cellcolor[rgb]{ .761,  .831,  .839}\textbf{0.50} & \textbf{0.83} & \cellcolor[rgb]{ .902,  .671,  .608}\textbf{0.04} & \cellcolor[rgb]{ .902,  .671,  .608}\textbf{0.03} & \cellcolor[rgb]{ .902,  .671,  .608}\textbf{0.05} \bigstrut[b]\\
\hline
\multirow{3}[2]{*}{\textbf{\makecell[l]{Default\\Sex}}} & \textbf{Original} & \textbf{0.20} & \textbf{0.81} & \textbf{0.03} & \textbf{0.03} & \textbf{0.05} & \textbf{0.15} & \textbf{0.80} & \textbf{0.02} & \textbf{0.02} & \textbf{0.04} & \textbf{0.73} & \textbf{0.67} & \textbf{0.17} & \textbf{0.14} & \textbf{0.12} & \textbf{0.73} & \textbf{0.67} & \textbf{0.17} & \textbf{0.14} & \textbf{0.12} & \textbf{0.36} & \textbf{0.81} & \textbf{0.03} & \textbf{0.02} & \textbf{0.02} & \textbf{0.33} & \textbf{0.82} & \textbf{0.03} & \textbf{0.02} & \textbf{0.03} \bigstrut[t]\\
      & \textbf{CoT-Phi} & \cellcolor[rgb]{ .761,  .831,  .839}\textbf{0.24} & \textbf{0.81} & \cellcolor[rgb]{ .902,  .671,  .608}\textbf{0.01} & \cellcolor[rgb]{ .902,  .671,  .608}\textbf{0.01} & \cellcolor[rgb]{ .902,  .671,  .608}\textbf{0.03} & \cellcolor[rgb]{ .761,  .831,  .839}\textbf{0.17} & \textbf{0.80} & \cellcolor[rgb]{ .902,  .671,  .608}\textbf{0.01} & \cellcolor[rgb]{ .902,  .671,  .608}\textbf{0.01} & \cellcolor[rgb]{ .902,  .671,  .608}\textbf{0.02} & \cellcolor[rgb]{ .761,  .831,  .839}\textbf{0.76} & \textbf{0.66} & \cellcolor[rgb]{ .902,  .671,  .608}\textbf{0.03} & \cellcolor[rgb]{ .902,  .671,  .608}\textbf{0.03} & \cellcolor[rgb]{ .902,  .671,  .608}\textbf{0.03} & \cellcolor[rgb]{ .761,  .831,  .839}\textbf{0.76} & \textbf{0.66} & \cellcolor[rgb]{ .902,  .671,  .608}\textbf{0.03} & \cellcolor[rgb]{ .902,  .671,  .608}\textbf{0.03} & \cellcolor[rgb]{ .902,  .671,  .608}\textbf{0.03} & \cellcolor[rgb]{ .761,  .831,  .839}\textbf{0.37} & \textbf{0.81} & \cellcolor[rgb]{ .902,  .671,  .608}\textbf{0.02} & \cellcolor[rgb]{ .902,  .671,  .608}\textbf{0.01} & \cellcolor[rgb]{ .902,  .671,  .608}\textbf{0.02} & \cellcolor[rgb]{ .761,  .831,  .839}\textbf{0.36} & \textbf{0.82} & \cellcolor[rgb]{ .902,  .671,  .608}\textbf{0.02} & \cellcolor[rgb]{ .902,  .671,  .608}\textbf{0.01} & \cellcolor[rgb]{ .902,  .671,  .608}\textbf{0.02} \\
      & \textbf{CoT-Opt} & \cellcolor[rgb]{ .761,  .831,  .839}\textbf{0.22} & \textbf{0.81} & \cellcolor[rgb]{ .902,  .671,  .608}\textbf{0.02} & \cellcolor[rgb]{ .902,  .671,  .608}\textbf{0.02} & \cellcolor[rgb]{ .902,  .671,  .608}\textbf{0.03} & \cellcolor[rgb]{ .761,  .831,  .839}\textbf{0.16} & \textbf{0.80} & \cellcolor[rgb]{ .902,  .671,  .608}\textbf{0.01} & \cellcolor[rgb]{ .902,  .671,  .608}\textbf{0.01} & \cellcolor[rgb]{ .902,  .671,  .608}\textbf{0.02} & \cellcolor[rgb]{ .761,  .831,  .839}\textbf{0.75} & \textbf{0.66} & \cellcolor[rgb]{ .902,  .671,  .608}\textbf{0.03} & \cellcolor[rgb]{ .902,  .671,  .608}\textbf{0.03} & \cellcolor[rgb]{ .902,  .671,  .608}\textbf{0.04} & \cellcolor[rgb]{ .761,  .831,  .839}\textbf{0.75} & \textbf{0.66} & \cellcolor[rgb]{ .902,  .671,  .608}\textbf{0.04} & \cellcolor[rgb]{ .902,  .671,  .608}\textbf{0.03} & \cellcolor[rgb]{ .902,  .671,  .608}\textbf{0.03} & \cellcolor[rgb]{ .761,  .831,  .839}\textbf{0.36} & \textbf{0.81} & \cellcolor[rgb]{ .902,  .671,  .608}\textbf{0.02} & \cellcolor[rgb]{ .902,  .671,  .608}\textbf{0.02} & \cellcolor[rgb]{ .902,  .671,  .608}\textbf{0.02} & \cellcolor[rgb]{ .761,  .831,  .839}\textbf{0.35} & \textbf{0.82} & \cellcolor[rgb]{ .902,  .671,  .608}\textbf{0.02} & \cellcolor[rgb]{ .902,  .671,  .608}\textbf{0.02} & \cellcolor[rgb]{ .902,  .671,  .608}\textbf{0.03} \bigstrut[b]\\
\hline
\multirow{3}[2]{*}{\textbf{\makecell[l]{Default\\Age}}} & \textbf{Original} & \textbf{0.22} & \textbf{0.81} & \textbf{0.04} & \textbf{0.04} & \textbf{0.06} & \textbf{0.16} & \textbf{0.80} & \textbf{0.02} & \textbf{0.02} & \textbf{0.03} & \textbf{0.71} & \textbf{0.67} & \textbf{0.17} & \textbf{0.14} & \textbf{0.13} & \textbf{0.71} & \textbf{0.67} & \textbf{0.17} & \textbf{0.14} & \textbf{0.12} & \textbf{0.36} & \textbf{0.81} & \textbf{0.05} & \textbf{0.05} & \textbf{0.07} & \textbf{0.35} & \textbf{0.82} & \textbf{0.06} & \textbf{0.06} & \textbf{0.08} \bigstrut[t]\\
      & \textbf{CoT-Phi} & \cellcolor[rgb]{ .761,  .831,  .839}\textbf{0.23} & \textbf{0.81} & \cellcolor[rgb]{ .902,  .671,  .608}\textbf{0.01} & \cellcolor[rgb]{ .902,  .671,  .608}\textbf{0.04} & \cellcolor[rgb]{ .902,  .671,  .608}\textbf{0.07} & \cellcolor[rgb]{ .761,  .831,  .839}\textbf{0.17} & \textbf{0.80} & \cellcolor[rgb]{ .902,  .671,  .608}\textbf{0.01} & \cellcolor[rgb]{ .902,  .671,  .608}\textbf{0.04} & \cellcolor[rgb]{ .902,  .671,  .608}\textbf{0.07} & \cellcolor[rgb]{ .761,  .831,  .839}\textbf{0.74} & \textbf{0.66} & \cellcolor[rgb]{ .902,  .671,  .608}\textbf{0.07} & \cellcolor[rgb]{ .902,  .671,  .608}\textbf{0.04} & \cellcolor[rgb]{ .902,  .671,  .608}\textbf{0.04} & \cellcolor[rgb]{ .761,  .831,  .839}\textbf{0.74} & \textbf{0.67} & \cellcolor[rgb]{ .902,  .671,  .608}\textbf{0.08} & \cellcolor[rgb]{ .902,  .671,  .608}\textbf{0.05} & \cellcolor[rgb]{ .902,  .671,  .608}\textbf{0.04} & \cellcolor[rgb]{ .761,  .831,  .839}\textbf{0.36} & \textbf{0.81} & \cellcolor[rgb]{ .902,  .671,  .608}\textbf{0.04} & \cellcolor[rgb]{ .902,  .671,  .608}\textbf{0.03} & \cellcolor[rgb]{ .902,  .671,  .608}\textbf{0.04} & \cellcolor[rgb]{ .761,  .831,  .839}\textbf{0.35} & \textbf{0.81} & \cellcolor[rgb]{ .902,  .671,  .608}\textbf{0.03} & \cellcolor[rgb]{ .902,  .671,  .608}\textbf{0.03} & \cellcolor[rgb]{ .902,  .671,  .608}\textbf{0.04} \\
      & \textbf{CoT-Opt} & \cellcolor[rgb]{ .761,  .831,  .839}\textbf{0.23} & \textbf{0.81} & \cellcolor[rgb]{ .902,  .671,  .608}\textbf{0.01} & \cellcolor[rgb]{ .902,  .671,  .608}\textbf{0.02} & \cellcolor[rgb]{ .902,  .671,  .608}\textbf{0.03} & \cellcolor[rgb]{ .761,  .831,  .839}\textbf{0.17} & \textbf{0.80} & \cellcolor[rgb]{ .902,  .671,  .608}\textbf{0.01} & \cellcolor[rgb]{ .902,  .671,  .608}\textbf{0.02} & \cellcolor[rgb]{ .902,  .671,  .608}\textbf{0.04} & \cellcolor[rgb]{ .761,  .831,  .839}\textbf{0.73} & \textbf{0.66} & \cellcolor[rgb]{ .902,  .671,  .608}\textbf{0.08} & \cellcolor[rgb]{ .902,  .671,  .608}\textbf{0.06} & \cellcolor[rgb]{ .902,  .671,  .608}\textbf{0.05} & \cellcolor[rgb]{ .761,  .831,  .839}\textbf{0.73} & \textbf{0.66} & \cellcolor[rgb]{ .902,  .671,  .608}\textbf{0.07} & \cellcolor[rgb]{ .902,  .671,  .608}\textbf{0.05} & \cellcolor[rgb]{ .902,  .671,  .608}\textbf{0.04} & \cellcolor[rgb]{ .761,  .831,  .839}\textbf{0.36} & \textbf{0.81} & \cellcolor[rgb]{ .902,  .671,  .608}\textbf{0.04} & \cellcolor[rgb]{ .902,  .671,  .608}\textbf{0.03} & \cellcolor[rgb]{ .902,  .671,  .608}\textbf{0.05} & \cellcolor[rgb]{ .761,  .831,  .839}\textbf{0.35} & \textbf{0.82} & \cellcolor[rgb]{ .902,  .671,  .608}\textbf{0.04} & \cellcolor[rgb]{ .902,  .671,  .608}\textbf{0.03} & \cellcolor[rgb]{ .902,  .671,  .608}\textbf{0.05} \bigstrut[b]\\
\hline
\multirow{3}[2]{*}{\textbf{\makecell[l]{Compas\\Sex}}} & \textbf{Original} & \textbf{0.72} & \textbf{0.68} & \textbf{0.23} & \textbf{0.20} & \textbf{0.16} & \textbf{0.72} & \textbf{0.68} & \textbf{0.22} & \textbf{0.19} & \textbf{0.15} & \textbf{0.35} & \textbf{0.81} & \textbf{0.03} & \textbf{0.02} & \textbf{0.03} & \textbf{0.36} & \textbf{0.82} & \textbf{0.03} & \textbf{0.02} & \textbf{0.02} & \textbf{0.70} & \textbf{0.65} & \textbf{0.13} & \textbf{0.11} & \textbf{0.09} & \textbf{0.73} & \textbf{0.68} & \textbf{0.21} & \textbf{0.18} & \textbf{0.15} \bigstrut[t]\\
      & \textbf{CoT-Phi} & \cellcolor[rgb]{ .761,  .831,  .839}\textbf{0.75} & \textbf{0.68} & \cellcolor[rgb]{ .902,  .671,  .608}\textbf{0.06} & \cellcolor[rgb]{ .902,  .671,  .608}\textbf{0.04} & \cellcolor[rgb]{ .902,  .671,  .608}\textbf{0.03} & \cellcolor[rgb]{ .761,  .831,  .839}\textbf{0.75} & \textbf{0.68} & \cellcolor[rgb]{ .902,  .671,  .608}\textbf{0.06} & \cellcolor[rgb]{ .902,  .671,  .608}\textbf{0.04} & \cellcolor[rgb]{ .902,  .671,  .608}\textbf{0.03} & \cellcolor[rgb]{ .761,  .831,  .839}\textbf{0.37} & \textbf{0.81} & \cellcolor[rgb]{ .902,  .671,  .608}\textbf{0.01} & \cellcolor[rgb]{ .902,  .671,  .608}\textbf{0.01} & \cellcolor[rgb]{ .902,  .671,  .608}\textbf{0.02} & \cellcolor[rgb]{ .761,  .831,  .839}\textbf{0.37} & \textbf{0.82} & \cellcolor[rgb]{ .902,  .671,  .608}\textbf{0.02} & \cellcolor[rgb]{ .902,  .671,  .608}\textbf{0.01} & \cellcolor[rgb]{ .902,  .671,  .608}\textbf{0.01} & \cellcolor[rgb]{ .761,  .831,  .839}\textbf{0.72} & \textbf{0.64} & \cellcolor[rgb]{ .902,  .671,  .608}\textbf{0.04} & \cellcolor[rgb]{ .902,  .671,  .608}\textbf{0.04} & \cellcolor[rgb]{ .902,  .671,  .608}\textbf{0.05} & \cellcolor[rgb]{ .761,  .831,  .839}\textbf{0.76} & \textbf{0.68} & \cellcolor[rgb]{ .902,  .671,  .608}\textbf{0.05} & \cellcolor[rgb]{ .902,  .671,  .608}\textbf{0.03} & \cellcolor[rgb]{ .902,  .671,  .608}\textbf{0.03} \\
      & \textbf{CoT-Opt} & \cellcolor[rgb]{ .761,  .831,  .839}\textbf{0.75} & \textbf{0.68} & \cellcolor[rgb]{ .902,  .671,  .608}\textbf{0.05} & \cellcolor[rgb]{ .902,  .671,  .608}\textbf{0.03} & \cellcolor[rgb]{ .902,  .671,  .608}\textbf{0.03} & \cellcolor[rgb]{ .761,  .831,  .839}\textbf{0.75} & \textbf{0.68} & \cellcolor[rgb]{ .902,  .671,  .608}\textbf{0.05} & \cellcolor[rgb]{ .902,  .671,  .608}\textbf{0.03} & \cellcolor[rgb]{ .902,  .671,  .608}\textbf{0.03} & \cellcolor[rgb]{ .761,  .831,  .839}\textbf{0.35} & \textbf{0.81} & \cellcolor[rgb]{ .902,  .671,  .608}\textbf{0.03} & \cellcolor[rgb]{ .902,  .671,  .608}\textbf{0.02} & \cellcolor[rgb]{ .902,  .671,  .608}\textbf{0.02} & \cellcolor[rgb]{ .761,  .831,  .839}\textbf{0.36} & \textbf{0.82} & \cellcolor[rgb]{ .902,  .671,  .608}\textbf{0.03} & \cellcolor[rgb]{ .902,  .671,  .608}\textbf{0.02} & \cellcolor[rgb]{ .902,  .671,  .608}\textbf{0.02} & \cellcolor[rgb]{ .761,  .831,  .839}\textbf{0.72} & \textbf{0.64} & \cellcolor[rgb]{ .902,  .671,  .608}\textbf{0.04} & \cellcolor[rgb]{ .902,  .671,  .608}\textbf{0.05} & \cellcolor[rgb]{ .902,  .671,  .608}\textbf{0.05} & \cellcolor[rgb]{ .761,  .831,  .839}\textbf{0.76} & \textbf{0.68} & \cellcolor[rgb]{ .902,  .671,  .608}\textbf{0.05} & \cellcolor[rgb]{ .902,  .671,  .608}\textbf{0.03} & \cellcolor[rgb]{ .902,  .671,  .608}\textbf{0.03} \bigstrut[b]\\
\hline
\multirow{3}[2]{*}{\textbf{\makecell[l]{Compas\\Race}}} & \textbf{Original} & \textbf{0.71} & \textbf{0.68} & \textbf{0.18} & \textbf{0.15} & \textbf{0.12} & \textbf{0.71} & \textbf{0.68} & \textbf{0.18} & \textbf{0.15} & \textbf{0.12} & \textbf{0.35} & \textbf{0.81} & \textbf{0.06} & \textbf{0.05} & \textbf{0.07} & \textbf{0.36} & \textbf{0.82} & \textbf{0.06} & \textbf{0.06} & \textbf{0.09} & \textbf{0.68} & \textbf{0.65} & \textbf{0.14} & \textbf{0.11} & \textbf{0.10} & \textbf{0.72} & \textbf{0.68} & \textbf{0.18} & \textbf{0.15} & \textbf{0.13} \bigstrut[t]\\
      & \textbf{CoT-Phi} & \cellcolor[rgb]{ .761,  .831,  .839}\textbf{0.74} & \textbf{0.67} & \cellcolor[rgb]{ .902,  .671,  .608}\textbf{0.06} & \cellcolor[rgb]{ .902,  .671,  .608}\textbf{0.03} & \cellcolor[rgb]{ .902,  .671,  .608}\textbf{0.02} & \cellcolor[rgb]{ .761,  .831,  .839}\textbf{0.74} & \textbf{0.68} & \cellcolor[rgb]{ .902,  .671,  .608}\textbf{0.06} & \cellcolor[rgb]{ .902,  .671,  .608}\textbf{0.04} & \cellcolor[rgb]{ .902,  .671,  .608}\textbf{0.02} & \cellcolor[rgb]{ .761,  .831,  .839}\textbf{0.36} & \textbf{0.81} & \cellcolor[rgb]{ .902,  .671,  .608}\textbf{0.03} & \cellcolor[rgb]{ .902,  .671,  .608}\textbf{0.02} & \cellcolor[rgb]{ .902,  .671,  .608}\textbf{0.02} & \cellcolor[rgb]{ .761,  .831,  .839}\textbf{0.36} & \textbf{0.82} & \cellcolor[rgb]{ .902,  .671,  .608}\textbf{0.04} & \cellcolor[rgb]{ .902,  .671,  .608}\textbf{0.03} & \cellcolor[rgb]{ .902,  .671,  .608}\textbf{0.04} & \cellcolor[rgb]{ .761,  .831,  .839}\textbf{0.71} & \textbf{0.64} & \cellcolor[rgb]{ .902,  .671,  .608}\textbf{0.05} & \cellcolor[rgb]{ .902,  .671,  .608}\textbf{0.03} & \cellcolor[rgb]{ .902,  .671,  .608}\textbf{0.03} & \cellcolor[rgb]{ .761,  .831,  .839}\textbf{0.74} & \textbf{0.67} & \cellcolor[rgb]{ .902,  .671,  .608}\textbf{0.06} & \cellcolor[rgb]{ .902,  .671,  .608}\textbf{0.04} & \cellcolor[rgb]{ .902,  .671,  .608}\textbf{0.04} \\
      & \textbf{CoT-Opt} & \cellcolor[rgb]{ .761,  .831,  .839}\textbf{0.74} & \textbf{0.68} & \cellcolor[rgb]{ .902,  .671,  .608}\textbf{0.06} & \cellcolor[rgb]{ .902,  .671,  .608}\textbf{0.03} & \cellcolor[rgb]{ .902,  .671,  .608}\textbf{0.02} & \cellcolor[rgb]{ .761,  .831,  .839}\textbf{0.74} & \textbf{0.68} & \cellcolor[rgb]{ .902,  .671,  .608}\textbf{0.06} & \cellcolor[rgb]{ .902,  .671,  .608}\textbf{0.04} & \cellcolor[rgb]{ .902,  .671,  .608}\textbf{0.02} & \cellcolor[rgb]{ .761,  .831,  .839}\textbf{0.36} & \textbf{0.81} & \cellcolor[rgb]{ .902,  .671,  .608}\textbf{0.04} & \cellcolor[rgb]{ .902,  .671,  .608}\textbf{0.03} & \cellcolor[rgb]{ .902,  .671,  .608}\textbf{0.04} & \cellcolor[rgb]{ .761,  .831,  .839}\textbf{0.36} & \textbf{0.82} & \cellcolor[rgb]{ .902,  .671,  .608}\textbf{0.05} & \cellcolor[rgb]{ .902,  .671,  .608}\textbf{0.04} & \cellcolor[rgb]{ .902,  .671,  .608}\textbf{0.06} & \cellcolor[rgb]{ .761,  .831,  .839}\textbf{0.71} & \textbf{0.64} & \cellcolor[rgb]{ .902,  .671,  .608}\textbf{0.05} & \cellcolor[rgb]{ .902,  .671,  .608}\textbf{0.03} & \cellcolor[rgb]{ .902,  .671,  .608}\textbf{0.03} & \cellcolor[rgb]{ .761,  .831,  .839}\textbf{0.73} & \textbf{0.67} & \cellcolor[rgb]{ .902,  .671,  .608}\textbf{0.06} & \cellcolor[rgb]{ .902,  .671,  .608}\textbf{0.04} & \cellcolor[rgb]{ .902,  .671,  .608}\textbf{0.04} \bigstrut[b]\\
\hline
\multirow{3}[2]{*}{\textbf{\makecell[l]{Mep1\\Sex}}} & \textbf{Original} & \textbf{0.32} & \textbf{0.86} & \textbf{0.06} & \textbf{0.05} & \textbf{0.07} & \textbf{0.32} & \textbf{0.86} & \textbf{0.05} & \textbf{0.03} & \textbf{0.04} & \textbf{0.41} & \textbf{0.86} & \textbf{0.06} & \textbf{0.04} & \textbf{0.04} & \textbf{0.41} & \textbf{0.86} & \textbf{0.06} & \textbf{0.03} & \textbf{0.04} & \textbf{0.38} & \textbf{0.86} & \textbf{0.04} & \textbf{0.02} & \textbf{0.03} & \textbf{0.35} & \textbf{0.86} & \textbf{0.07} & \textbf{0.05} & \textbf{0.07} \bigstrut[t]\\
      & \textbf{CoT-Phi} & \cellcolor[rgb]{ .761,  .831,  .839}\textbf{0.39} & \textbf{0.86} & \cellcolor[rgb]{ .902,  .671,  .608}\textbf{0.02} & \cellcolor[rgb]{ .902,  .671,  .608}\textbf{0.03} & \cellcolor[rgb]{ .902,  .671,  .608}\textbf{0.07} & \cellcolor[rgb]{ .761,  .831,  .839}\textbf{0.37} & \textbf{0.86} & \cellcolor[rgb]{ .902,  .671,  .608}\textbf{0.02} & \cellcolor[rgb]{ .902,  .671,  .608}\textbf{0.02} & \cellcolor[rgb]{ .902,  .671,  .608}\textbf{0.05} & \cellcolor[rgb]{ .761,  .831,  .839}\textbf{0.44} & \textbf{0.86} & \cellcolor[rgb]{ .902,  .671,  .608}\textbf{0.03} & \cellcolor[rgb]{ .902,  .671,  .608}\textbf{0.02} & \cellcolor[rgb]{ .902,  .671,  .608}\textbf{0.05} & \cellcolor[rgb]{ .761,  .831,  .839}\textbf{0.44} & \textbf{0.86} & \cellcolor[rgb]{ .902,  .671,  .608}\textbf{0.03} & \cellcolor[rgb]{ .902,  .671,  .608}\textbf{0.02} & \cellcolor[rgb]{ .902,  .671,  .608}\textbf{0.04} & \cellcolor[rgb]{ .761,  .831,  .839}\textbf{0.41} & \textbf{0.86} & \cellcolor[rgb]{ .902,  .671,  .608}\textbf{0.03} & \cellcolor[rgb]{ .902,  .671,  .608}\textbf{0.02} & \cellcolor[rgb]{ .902,  .671,  .608}\textbf{0.05} & \cellcolor[rgb]{ .761,  .831,  .839}\textbf{0.44} & \textbf{0.86} & \cellcolor[rgb]{ .902,  .671,  .608}\textbf{0.02} & \cellcolor[rgb]{ .902,  .671,  .608}\textbf{0.04} & \cellcolor[rgb]{ .902,  .671,  .608}\textbf{0.07} \\
      & \textbf{CoT-Opt} & \cellcolor[rgb]{ .761,  .831,  .839}\textbf{0.36} & \textbf{0.86} & \cellcolor[rgb]{ .902,  .671,  .608}\textbf{0.04} & \cellcolor[rgb]{ .902,  .671,  .608}\textbf{0.02} & \cellcolor[rgb]{ .902,  .671,  .608}\textbf{0.03} & \cellcolor[rgb]{ .761,  .831,  .839}\textbf{0.35} & \textbf{0.86} & \cellcolor[rgb]{ .902,  .671,  .608}\textbf{0.04} & \cellcolor[rgb]{ .902,  .671,  .608}\textbf{0.01} & \cellcolor[rgb]{ .902,  .671,  .608}\textbf{0.02} & \cellcolor[rgb]{ .761,  .831,  .839}\textbf{0.43} & \textbf{0.85} & \cellcolor[rgb]{ .902,  .671,  .608}\textbf{0.04} & \cellcolor[rgb]{ .902,  .671,  .608}\textbf{0.02} & \cellcolor[rgb]{ .902,  .671,  .608}\textbf{0.03} & \cellcolor[rgb]{ .761,  .831,  .839}\textbf{0.43} & \textbf{0.86} & \cellcolor[rgb]{ .902,  .671,  .608}\textbf{0.04} & \cellcolor[rgb]{ .902,  .671,  .608}\textbf{0.02} & \cellcolor[rgb]{ .902,  .671,  .608}\textbf{0.03} & \cellcolor[rgb]{ .761,  .831,  .839}\textbf{0.39} & \textbf{0.86} & \cellcolor[rgb]{ .902,  .671,  .608}\textbf{0.03} & \cellcolor[rgb]{ .902,  .671,  .608}\textbf{0.01} & \cellcolor[rgb]{ .902,  .671,  .608}\textbf{0.03} & \cellcolor[rgb]{ .761,  .831,  .839}\textbf{0.39} & \textbf{0.86} & \cellcolor[rgb]{ .902,  .671,  .608}\textbf{0.04} & \cellcolor[rgb]{ .902,  .671,  .608}\textbf{0.02} & \cellcolor[rgb]{ .902,  .671,  .608}\textbf{0.03} \bigstrut[b]\\
\hline
\multirow{3}[2]{*}{\textbf{\makecell[l]{Mep1\\Race}}} & \textbf{Original} & \textbf{0.32} & \textbf{0.86} & \textbf{0.04} & \textbf{0.04} & \textbf{0.06} & \textbf{0.33} & \textbf{0.86} & \textbf{0.03} & \textbf{0.02} & \textbf{0.04} & \textbf{0.37} & \textbf{0.86} & \textbf{0.04} & \textbf{0.05} & \textbf{0.08} & \textbf{0.38} & \textbf{0.86} & \textbf{0.04} & \textbf{0.05} & \textbf{0.08} & \textbf{0.36} & \textbf{0.86} & \textbf{0.03} & \textbf{0.03} & \textbf{0.05} & \textbf{0.33} & \textbf{0.86} & \textbf{0.04} & \textbf{0.05} & \textbf{0.08} \bigstrut[t]\\
      & \textbf{CoT-Phi} & \cellcolor[rgb]{ .761,  .831,  .839}\textbf{0.36} & \textbf{0.86} & \cellcolor[rgb]{ .902,  .671,  .608}\textbf{0.02} & \cellcolor[rgb]{ .902,  .671,  .608}\textbf{0.02} & \cellcolor[rgb]{ .902,  .671,  .608}\textbf{0.03} & \cellcolor[rgb]{ .761,  .831,  .839}\textbf{0.35} & \textbf{0.86} & \cellcolor[rgb]{ .902,  .671,  .608}\textbf{0.01} & \cellcolor[rgb]{ .902,  .671,  .608}\textbf{0.02} & \cellcolor[rgb]{ .902,  .671,  .608}\textbf{0.03} & \cellcolor[rgb]{ .761,  .831,  .839}\textbf{0.41} & \textbf{0.86} & \cellcolor[rgb]{ .902,  .671,  .608}\textbf{0.02} & \cellcolor[rgb]{ .902,  .671,  .608}\textbf{0.02} & \cellcolor[rgb]{ .902,  .671,  .608}\textbf{0.04} & \cellcolor[rgb]{ .761,  .831,  .839}\textbf{0.41} & \textbf{0.86} & \cellcolor[rgb]{ .902,  .671,  .608}\textbf{0.02} & \cellcolor[rgb]{ .902,  .671,  .608}\textbf{0.02} & \cellcolor[rgb]{ .902,  .671,  .608}\textbf{0.04} & \cellcolor[rgb]{ .761,  .831,  .839}\textbf{0.37} & \textbf{0.86} & \cellcolor[rgb]{ .902,  .671,  .608}\textbf{0.02} & \cellcolor[rgb]{ .902,  .671,  .608}\textbf{0.02} & \cellcolor[rgb]{ .902,  .671,  .608}\textbf{0.03} & \cellcolor[rgb]{ .761,  .831,  .839}\textbf{0.37} & \textbf{0.86} & \cellcolor[rgb]{ .902,  .671,  .608}\textbf{0.02} & \cellcolor[rgb]{ .902,  .671,  .608}\textbf{0.02} & \cellcolor[rgb]{ .902,  .671,  .608}\textbf{0.03} \\
      & \textbf{CoT-Opt} & \cellcolor[rgb]{ .761,  .831,  .839}\textbf{0.37} & \textbf{0.86} & \cellcolor[rgb]{ .902,  .671,  .608}\textbf{0.01} & \cellcolor[rgb]{ .902,  .671,  .608}\textbf{0.02} & \cellcolor[rgb]{ .902,  .671,  .608}\textbf{0.04} & \cellcolor[rgb]{ .761,  .831,  .839}\textbf{0.35} & \textbf{0.86} & \cellcolor[rgb]{ .902,  .671,  .608}\textbf{0.01} & \cellcolor[rgb]{ .902,  .671,  .608}\textbf{0.02} & \cellcolor[rgb]{ .902,  .671,  .608}\textbf{0.03} & \cellcolor[rgb]{ .761,  .831,  .839}\textbf{0.42} & \textbf{0.86} & \cellcolor[rgb]{ .902,  .671,  .608}\textbf{0.02} & \cellcolor[rgb]{ .902,  .671,  .608}\textbf{0.01} & \cellcolor[rgb]{ .902,  .671,  .608}\textbf{0.03} & \cellcolor[rgb]{ .761,  .831,  .839}\textbf{0.42} & \textbf{0.86} & \cellcolor[rgb]{ .902,  .671,  .608}\textbf{0.02} & \cellcolor[rgb]{ .902,  .671,  .608}\textbf{0.02} & \cellcolor[rgb]{ .902,  .671,  .608}\textbf{0.04} & \cellcolor[rgb]{ .761,  .831,  .839}\textbf{0.37} & \textbf{0.86} & \cellcolor[rgb]{ .902,  .671,  .608}\textbf{0.03} & \cellcolor[rgb]{ .902,  .671,  .608}\textbf{0.02} & \cellcolor[rgb]{ .902,  .671,  .608}\textbf{0.04} & \cellcolor[rgb]{ .761,  .831,  .839}\textbf{0.36} & \textbf{0.86} & \cellcolor[rgb]{ .902,  .671,  .608}\textbf{0.02} & \cellcolor[rgb]{ .902,  .671,  .608}\textbf{0.02} & \cellcolor[rgb]{ .902,  .671,  .608}\textbf{0.03} \bigstrut[b]\\
\hline
\multirow{3}[2]{*}{\textbf{\makecell[l]{Mep2\\Sex}}} & \textbf{Original} & \textbf{0.26} & \textbf{0.86} & \textbf{0.06} & \textbf{0.05} & \textbf{0.08} & \textbf{0.27} & \textbf{0.86} & \textbf{0.04} & \textbf{0.03} & \textbf{0.04} & \textbf{0.34} & \textbf{0.85} & \textbf{0.07} & \textbf{0.05} & \textbf{0.06} & \textbf{0.33} & \textbf{0.86} & \textbf{0.06} & \textbf{0.04} & \textbf{0.06} & \textbf{0.32} & \textbf{0.86} & \textbf{0.04} & \textbf{0.02} & \textbf{0.02} & \textbf{0.29} & \textbf{0.86} & \textbf{0.06} & \textbf{0.05} & \textbf{0.07} \bigstrut[t]\\
      & \textbf{CoT-Phi} & \cellcolor[rgb]{ .761,  .831,  .839}\textbf{0.34} & \textbf{0.86} & \cellcolor[rgb]{ .902,  .671,  .608}\textbf{0.02} & \cellcolor[rgb]{ .902,  .671,  .608}\textbf{0.03} & \cellcolor[rgb]{ .902,  .671,  .608}\textbf{0.05} & \cellcolor[rgb]{ .761,  .831,  .839}\textbf{0.32} & \textbf{0.86} & \cellcolor[rgb]{ .902,  .671,  .608}\textbf{0.02} & \cellcolor[rgb]{ .902,  .671,  .608}\textbf{0.02} & \cellcolor[rgb]{ .902,  .671,  .608}\textbf{0.05} & \cellcolor[rgb]{ .761,  .831,  .839}\textbf{0.39} & \textbf{0.85} & \cellcolor[rgb]{ .902,  .671,  .608}\textbf{0.03} & \cellcolor[rgb]{ .902,  .671,  .608}\textbf{0.02} & \cellcolor[rgb]{ .902,  .671,  .608}\textbf{0.05} & \cellcolor[rgb]{ .761,  .831,  .839}\textbf{0.38} & \textbf{0.85} & \cellcolor[rgb]{ .902,  .671,  .608}\textbf{0.03} & \cellcolor[rgb]{ .902,  .671,  .608}\textbf{0.02} & \cellcolor[rgb]{ .902,  .671,  .608}\textbf{0.04} & \cellcolor[rgb]{ .761,  .831,  .839}\textbf{0.34} & \textbf{0.86} & \cellcolor[rgb]{ .902,  .671,  .608}\textbf{0.02} & \cellcolor[rgb]{ .902,  .671,  .608}\textbf{0.02} & \cellcolor[rgb]{ .902,  .671,  .608}\textbf{0.04} & \cellcolor[rgb]{ .761,  .831,  .839}\textbf{0.36} & \textbf{0.85} & \cellcolor[rgb]{ .902,  .671,  .608}\textbf{0.01} & \cellcolor[rgb]{ .902,  .671,  .608}\textbf{0.04} & \cellcolor[rgb]{ .902,  .671,  .608}\textbf{0.08} \\
      & \textbf{CoT-Opt} & \cellcolor[rgb]{ .761,  .831,  .839}\textbf{0.31} & \textbf{0.86} & \cellcolor[rgb]{ .902,  .671,  .608}\textbf{0.03} & \cellcolor[rgb]{ .902,  .671,  .608}\textbf{0.02} & \cellcolor[rgb]{ .902,  .671,  .608}\textbf{0.03} & \cellcolor[rgb]{ .761,  .831,  .839}\textbf{0.30} & \textbf{0.86} & \cellcolor[rgb]{ .902,  .671,  .608}\textbf{0.03} & \cellcolor[rgb]{ .902,  .671,  .608}\textbf{0.01} & \cellcolor[rgb]{ .902,  .671,  .608}\textbf{0.03} & \cellcolor[rgb]{ .761,  .831,  .839}\textbf{0.37} & \textbf{0.85} & \cellcolor[rgb]{ .902,  .671,  .608}\textbf{0.04} & \cellcolor[rgb]{ .902,  .671,  .608}\textbf{0.02} & \cellcolor[rgb]{ .902,  .671,  .608}\textbf{0.03} & \cellcolor[rgb]{ .761,  .831,  .839}\textbf{0.37} & \textbf{0.86} & \cellcolor[rgb]{ .902,  .671,  .608}\textbf{0.04} & \cellcolor[rgb]{ .902,  .671,  .608}\textbf{0.02} & \cellcolor[rgb]{ .902,  .671,  .608}\textbf{0.03} & \cellcolor[rgb]{ .761,  .831,  .839}\textbf{0.34} & \textbf{0.86} & \cellcolor[rgb]{ .902,  .671,  .608}\textbf{0.03} & \cellcolor[rgb]{ .902,  .671,  .608}\textbf{0.01} & \cellcolor[rgb]{ .902,  .671,  .608}\textbf{0.03} & \cellcolor[rgb]{ .761,  .831,  .839}\textbf{0.36} & \textbf{0.85} & \cellcolor[rgb]{ .902,  .671,  .608}\textbf{0.03} & \cellcolor[rgb]{ .902,  .671,  .608}\textbf{0.02} & \cellcolor[rgb]{ .902,  .671,  .608}\textbf{0.04} \bigstrut[b]\\
\hline
\multirow{3}[2]{*}{\textbf{\makecell[l]{Mep2\\Race}}} & \textbf{Original} & \textbf{0.25} & \textbf{0.86} & \textbf{0.04} & \textbf{0.05} & \textbf{0.08} & \textbf{0.27} & \textbf{0.86} & \textbf{0.03} & \textbf{0.02} & \textbf{0.04} & \textbf{0.32} & \textbf{0.85} & \textbf{0.05} & \textbf{0.05} & \textbf{0.07} & \textbf{0.33} & \textbf{0.86} & \textbf{0.05} & \textbf{0.04} & \textbf{0.06} & \textbf{0.31} & \textbf{0.86} & \textbf{0.03} & \textbf{0.02} & \textbf{0.03} & \textbf{0.28} & \textbf{0.86} & \textbf{0.05} & \textbf{0.05} & \textbf{0.08} \bigstrut[t]\\
      & \textbf{CoT-Phi} & \cellcolor[rgb]{ .761,  .831,  .839}\textbf{0.32} & \textbf{0.86} & \cellcolor[rgb]{ .902,  .671,  .608}\textbf{0.02} & \cellcolor[rgb]{ .902,  .671,  .608}\textbf{0.01} & \cellcolor[rgb]{ .902,  .671,  .608}\textbf{0.03} & \cellcolor[rgb]{ .761,  .831,  .839}\textbf{0.31} & \textbf{0.86} & \cellcolor[rgb]{ .902,  .671,  .608}\textbf{0.01} & \cellcolor[rgb]{ .902,  .671,  .608}\textbf{0.01} & \cellcolor[rgb]{ .902,  .671,  .608}\textbf{0.03} & \cellcolor[rgb]{ .761,  .831,  .839}\textbf{0.37} & \textbf{0.85} & \cellcolor[rgb]{ .902,  .671,  .608}\textbf{0.02} & \cellcolor[rgb]{ .902,  .671,  .608}\textbf{0.02} & \cellcolor[rgb]{ .902,  .671,  .608}\textbf{0.03} & \cellcolor[rgb]{ .761,  .831,  .839}\textbf{0.37} & \textbf{0.85} & \cellcolor[rgb]{ .902,  .671,  .608}\textbf{0.02} & \cellcolor[rgb]{ .902,  .671,  .608}\textbf{0.01} & \cellcolor[rgb]{ .902,  .671,  .608}\textbf{0.03} & \cellcolor[rgb]{ .761,  .831,  .839}\textbf{0.33} & \textbf{0.86} & \cellcolor[rgb]{ .902,  .671,  .608}\textbf{0.02} & \cellcolor[rgb]{ .902,  .671,  .608}\textbf{0.01} & \cellcolor[rgb]{ .902,  .671,  .608}\textbf{0.02} & \cellcolor[rgb]{ .761,  .831,  .839}\textbf{0.36} & \textbf{0.86} & \cellcolor[rgb]{ .902,  .671,  .608}\textbf{0.01} & \cellcolor[rgb]{ .902,  .671,  .608}\textbf{0.02} & \cellcolor[rgb]{ .902,  .671,  .608}\textbf{0.04} \\
      & \textbf{CoT-Opt} & \cellcolor[rgb]{ .761,  .831,  .839}\textbf{0.31} & \textbf{0.86} & \cellcolor[rgb]{ .902,  .671,  .608}\textbf{0.02} & \cellcolor[rgb]{ .902,  .671,  .608}\textbf{0.02} & \cellcolor[rgb]{ .902,  .671,  .608}\textbf{0.03} & \cellcolor[rgb]{ .761,  .831,  .839}\textbf{0.30} & \textbf{0.86} & \cellcolor[rgb]{ .902,  .671,  .608}\textbf{0.02} & \cellcolor[rgb]{ .902,  .671,  .608}\textbf{0.01} & \cellcolor[rgb]{ .902,  .671,  .608}\textbf{0.02} & \cellcolor[rgb]{ .761,  .831,  .839}\textbf{0.38} & \textbf{0.85} & \cellcolor[rgb]{ .902,  .671,  .608}\textbf{0.03} & \cellcolor[rgb]{ .902,  .671,  .608}\textbf{0.02} & \cellcolor[rgb]{ .902,  .671,  .608}\textbf{0.03} & \cellcolor[rgb]{ .761,  .831,  .839}\textbf{0.36} & \textbf{0.86} & \cellcolor[rgb]{ .902,  .671,  .608}\textbf{0.02} & \cellcolor[rgb]{ .902,  .671,  .608}\textbf{0.01} & \cellcolor[rgb]{ .902,  .671,  .608}\textbf{0.03} & \cellcolor[rgb]{ .761,  .831,  .839}\textbf{0.32} & \textbf{0.86} & \cellcolor[rgb]{ .902,  .671,  .608}\textbf{0.02} & \cellcolor[rgb]{ .902,  .671,  .608}\textbf{0.01} & \cellcolor[rgb]{ .902,  .671,  .608}\textbf{0.02} & \cellcolor[rgb]{ .761,  .831,  .839}\textbf{0.35} & \textbf{0.86} & \cellcolor[rgb]{ .902,  .671,  .608}\textbf{0.02} & \cellcolor[rgb]{ .902,  .671,  .608}\textbf{0.02} & \cellcolor[rgb]{ .902,  .671,  .608}\textbf{0.03} \bigstrut[b]\\
\hline
\end{tabular}%

    }
  \label{tab:metrics_value}%
\end{table*}%

\subsubsection{Trade-off Baseline}
To measure the effectiveness of the trade-off between model fairness and performance, Hort et al. proposed Fairea \cite{hort2021fairea} at ESEC/FSE 2021, introducing a trade-off baseline using AOD-Accuracy and SPD-Accuracy metrics. To provide a more comprehensive evaluation, Chen et al. extended this baseline to fifteen fairness-performance metrics (combinations of three fairness metrics and five performance metrics) \cite{chen2022maat}. The baseline categorizes the effectiveness of the fairness-performance trade-off into five levels: ``win-win'' (improvement in both performance and fairness), ``good'' (improved fairness, reduced performance, but still surpassing the Fairea Baseline), ``inverted'' (improved performance but reduced fairness), ``poor'' (improved fairness, reduced performance, but not surpassing the Fairea Baseline), and ``lose-lose'' (reduction in both fairness and performance). In this work, we follow prior studies \cite{xiao2024mirrorfair, chen2024fairness, joshi2024fairgenerate} and adopt the Fairea Trade-off Baseline to evaluate existing methods and CoT.

\subsection{Statistical Analysis}
% We use the non-parametric Mann Whitney U-test [57] (which suits our purpose well as it does not assume normality) to test whether the fairness/performance is significantly improved/decreased. The fairness/performance change is considered statistically significant, only if the $\rho$-value of the computed statistic is lower than 0.05.

To enhance the reliability of our experimental results, we follow previous work \cite{chen2022maat, xiao2024mirrorfair} and recent empirical studies \cite{chen2024fairness, chen2023comprehensive}, employing two statistical tools: the Mann-Whitney U-test \cite{mann1947test} and Cliff's $\delta$ to analyze our raw results. In RQ2 and RQ4, we use the Mann-Whitney U-test to assess whether improvements in fairness after applying bias mitigation methods are statistically significant. Consistent with previous studies \cite{chen2022maat, xiao2024mirrorfair}, we consider improvements statistically significant when the $p$-value is below 0.05. Additionally, we employ Cliff's $\delta$, which is widely used in SE research \cite{bennin2017significant, vargha2000critique, chen2024fairness}, to measure the effect size of the impact. In line with the literature \cite{chen2024fairness}, we set our threshold at 0.428; when the absolute value of $\delta$ is no less than 0.428, we consider the change to have a large effect.

% To enhance the reliability of our experimental results, we follow previous work \cite{chen2022maat, xiao2024mirrorfair} and recent empirical study \cite{chen2024fairness, chen2023comprehensive}, and employ two statistical tools, the Mann-Whitney U-test \cite{mann1947test} and Cliff's $\delta$, to analyze the raw experimental results. In RQ2 and RQ4, we use the Mann-Whitney U-test to evaluate whether the improvement of fairness after applying bias mitigation methods is statistically significant. Our threshold aligns with previous work \cite{chen2022maat, xiao2024mirrorfair}, when the $p$-value is lower than 0.05, we identify the improvement as statistically significant. Additionally, we employ Cliff's $\delta$, which is widely adopted in SE research \cite{bennin2017significant, vargha2000critique, chen2024fairness} to measure the effect size of the impact. Also align with the literature \cite{chen2024fairness}, we set our threshold to be 0.428. When an absolute value of $\delta$ is no less than 0.428, we identify the change as large effect.

% \input{1-tab-AllMetrics}

\subsection{Experiment Settings}
\label{sec:experimental setting}

Given the prominence of fairness as a rapidly growing field across various research communities, we have rigorously structured our experiments to ensure their reliability. To maintain fairness in comparison, we align our experimental setups, including benchmarking datasets, classification algorithms, evaluation metrics, and experimental environment, with those of recent empirical studies that comprehensively explored existing bias-mitigating methods. The experimental classifiers include Logistic Regression (LR) \cite{wright1995logistic}, Support Vector Machine (SVM) \cite{noble2006support}, Random Forest (RF) \cite{biau2016random}, and deep neural networks (DNN) \cite{lecun2015deep}. \color{black} We also include two modern classifiers, XGBoost (XGB)~\cite{chen2016xgboost} and LightGBM (LGBM)~\cite{ke2017lightgbm}, to enhance the robustness and comprehensiveness of our evaluation. \color{black} 

Regarding the implementation of existing methods, we meticulously replicated them using the IBM AIF360 toolkit~\cite{bellamy2019ai}, the source code released by the original authors~\cite{li2022training, joshi2024fairgenerate}, and recent empirical investigations~\cite{chen2023comprehensive, chen2024fairness}.  For each task, we conducted 20 repeated experiments to reduce random error, and the mean values across all runs are reported as the final results. \color{black}We adopted the iteration count as the random seed to split each dataset into a 70\% training set and a 30\% testing set, following recent empirical studies~\cite{chen2024fairness} to enhance the soundness of our evaluation.\color{black}  All experiments involving ML and DL models were conducted on a CPU platform using Python 3.12.

\color{black}

\section{Results}
\label{sec:result}
In this section, we present our experimental results by answering the RQs. Considering space limitations, we primarily report the statistical analysis results in the paper. All raw results for model fairness and performance across each experimental scenario are available in the supplementary materials.

\color{black}
\subsection{RQ1: Impact of CoT on ML Software}
\label{RQ1}

We answer this RQ through a comprehensive comparison between two implementations of CoT (CoT-Phi and CoT-Opt) and the original model without any bias mitigation methods, across the ML and DL scenarios on ten tasks involving five datasets. Additionally, we specifically analyze the impact of CoT on the model’s ability to assign correct favorable labels to unprivileged groups.

Table \ref{tab:metrics_value} presents the accuracy, three fairness metrics, and the true positive rate (TPR) for unprivileged groups. Specifically, CoT, including both CoT-Phi and CoT-Opt, significantly affects the TPR of unprivileged groups, typically increasing its value. This indicates that the model's tendency to assign favorable labels to unprivileged groups is strengthened. \color{black}The positive impact of CoT on the unprivileged group does not substantially compromise overall model accuracy, while it significantly reduces bias as measured by the SPD, AOD, and EOD metrics in the ML software. In particular, CoT-Phi increases TPR-U by 17\% and decreases SPD, AOD, and EOD by 57\%, 55\%, and 41\%, respectively. CoT-Opt increases TPR-U by 15\%, and decreases SPD, AOD, and EOD by 47\%, 58\%, and 51\%, respectively. 
\color{black}

\finding{Both CoT-Phi and CoT-Opt significantly \textbf{enhance true positive rate for unprivileged groups} without substantially compromising overall accuracy, demonstrating strong effectiveness in \textbf{trade-off model performance} between different groups. \color{black} Additionally, this performance trade-off enables CoT to achieve significant improvements in model fairness. Specifically, CoT-Opt decreases the SPD, AOD, and EOD bias by 47\%, 58\%, and 51\%.\color{black}}

\subsection{RQ2: Effectiveness of in Improving Fairness}
\label{RQ2}
In RQ\ref{RQ1}, we examine the impact of CoT on model performance and fairness across scenarios six models and ten classification tasks. In this RQ, we employ the Mann-Whitney U-Test as a statistical analysis tool to provide an overall understanding of the effectiveness of CoT in mitigating data bias and enhancing model fairness. Additionally, we compare the two implementations of CoT with existing methods across 60 scenarios (10 tasks $\times$ 6 models) to highlight the advantages of CoT. 

Specifically, we evaluate both existing methods and CoT from two perspectives. First, we treat each combination of dataset, sensitive attribute, model, and fairness metric as a distinct scenario. We then measure the proportion of scenarios in which fairness is improved and the effect is significantly large, using the Mann-Whitney U-test and Cliff’s $\delta$. Second, we assess the degree of change in fairness metrics, considering both absolute and relative changes.

% Table generated by Excel2LaTeX from sheet 'statistical_increase'
\begin{table}[htbp]
  \centering
  \caption{(RQ2) Proportions of scenarios that improve fairness and have a significantly large effect. The top three values are highlighted with \colorbox[rgb]{.651, .651, .651}{gray background}. \color{black}CoT-Opt achieves the highest proportions, with a 97.8\% fairness improvement rate and 83.3\% exhibiting a significantly large effect.\color{black}}
  \resizebox{\linewidth}{!}{
% Table generated by Excel2LaTeX from sheet 'statistical_increase'
% Table generated by Excel2LaTeX from sheet 'RQ-2-1'
\begin{tabular}{l|rr}
\hline
Method & \multicolumn{1}{c}{Fairness Increase} & \multicolumn{1}{c}{Fairness Increase with Large Effect} \bigstrut\\
\hline
Fair-SMOTE & 63.3\% & 53.3\% \bigstrut[t]\\
FairGenerate & 76.7\% & 56.7\% \\
LTDD  & 61.1\% & 38.3\% \\
FairMask & 90.6\% & 67.8\% \\
MirrorFair & \cellcolor[rgb]{ .651,  .651,  .651}92.2\% & \cellcolor[rgb]{ .651,  .651,  .651}77.8\% \\
CoT-Phi & \cellcolor[rgb]{ .651,  .651,  .651}92.2\% & \cellcolor[rgb]{ .651,  .651,  .651}81.7\% \\
CoT-Opt & \cellcolor[rgb]{ .651,  .651,  .651}\textbf{97.8\%} & \cellcolor[rgb]{ .651,  .651,  .651}\textbf{83.3\%} \bigstrut[b]\\
\hline
\end{tabular}%

    }
  \label{tab:fairness_increase_proportion}%
\end{table}%

\color{black}
Table \ref{tab:fairness_increase_proportion} presents the proportion of scenarios in which fairness is improved by existing methods and CoT. Among the baselines, MirrorFair achieves the highest proportion for both fairness improvement (92.2\%) and fairness improvement with a significantly large effect (77.8\%). CoT-Opt attains fairness improvement proportions of 97.8\% and 83.3\%, respectively, surpassing the state-of-the-art by six and three percentage points.

\color{black}

% Table generated by Excel2LaTeX from sheet 'fairness_value_change'
\begin{table}[htbp]
  \centering
  \caption{(RQ2) Absolute and relative changes (in parentheses) in bias metrics achieved by existing methods and CoT. The top three values are highlighted with \colorbox[rgb]{.651, .651, .651}{gray background}.\color{black} CoT-Opt achieves the best overall effectiveness, reducing SPD, AOD, and EOD by 46.9\%, 58.1\%, and 51.0\%, respectively.\color{black}  Notably, the ranking of absolute and relative changes may differ, as the base bias values in each scenario can vary.}
  \resizebox{\linewidth}{!}{
% Table generated by Excel2LaTeX from sheet 'fairness_value_change'
% Table generated by Excel2LaTeX from sheet 'fairness_value_change'
% Table generated by Excel2LaTeX from sheet 'RQ2-2'
\begin{tabular}{l|rr|rr|rr}
\hline
Method & \multicolumn{2}{c|}{SPD} & \multicolumn{2}{c|}{AOD} & \multicolumn{2}{c}{EOD} \bigstrut\\
\hline
FairGenerate & -0.026 & (-10.6\%) & -0.037 & (-32.9\%) & -0.049 & \cellcolor[rgb]{ .651,  .651,  .651}(-43.4\%) \bigstrut[t]\\
LTDD  & -0.030 & (-29.2\%) & -0.022 & (-9.8\%) & -0.018 & (-3.1\%) \\
Fair-SMOTE & -0.017 & (-0.036) & -0.028 & (-12.9\%) & -0.039 & (-29.3\%) \\
FairMask & -0.023 & (-24.4\%) & -0.033 & (-41.2\%) & -0.045 & (-41.3\%) \\
MirrorFair & \cellcolor[rgb]{ .651,  .651,  .651}-0.046 & \cellcolor[rgb]{ .651,  .651,  .651}(-50.9\%) & \cellcolor[rgb]{ .651,  .651,  .651}-0.042 & \cellcolor[rgb]{ .651,  .651,  .651}(-51.9\%) & \cellcolor[rgb]{ .651,  .651,  .651}-0.050 & \cellcolor[rgb]{ .651,  .651,  .651}(-42.0\%) \\
CoT-Phi & \cellcolor[rgb]{ .651,  .651,  .651}\textbf{-0.054} & \cellcolor[rgb]{ .651,  .651,  .651}\textbf{(-57.4\%)} & \cellcolor[rgb]{ .651,  .651,  .651}-0.051 & \cellcolor[rgb]{ .651,  .651,  .651}(-54.6\%) & \cellcolor[rgb]{ .651,  .651,  .651}-0.054 & (-40.6\%) \\
CoT-Opt & \cellcolor[rgb]{ .651,  .651,  .651}-0.049 & \cellcolor[rgb]{ .651,  .651,  .651}(-46.9\%) & \cellcolor[rgb]{ .651,  .651,  .651}\textbf{-0.052} & \cellcolor[rgb]{ .651,  .651,  .651}\textbf{(-58.1\%)} & \cellcolor[rgb]{ .651,  .651,  .651}\textbf{-0.060} & \cellcolor[rgb]{ .651,  .651,  .651}\textbf{(-51.0\%)} \bigstrut[b]\\
\hline
\end{tabular}%

    }
  \label{tab:fairness_metrics_change}%
\end{table}%

Table \ref{tab:fairness_metrics_change} presents the absolute and relative changes in three fairness metrics for five existing methods and two implementations of CoT. Different methods demonstrate varying strengths in mitigating bias as measured by the different metrics. \color{black} Among the baselines, MirrorFair achieves the greatest reduction in SPD bias by 0.046 (50.9\%), and FairGenerate achieves the largest reduction in EOD bias by 0.049 (43.4\%), while MirrorFair demonstrates the highest overall effectiveness across all three types of bias. For CoT, CoT-Phi achieves the greatest reduction in SPD by 0.054 (57.4\%), while CoT-Opt achieves the largest reductions in AOD and EOD by 0.052 (58.1\%) and 0.060 (51.0\%), respectively.

\color{black}

\finding{CoT outperforms the state-of-the-art in both the proportion of fairness improvement scenarios and the degree of changes in fairness metrics. \color{black}Specifically, CoT-Opt improves fairness in 97.8\% of scenarios, with 83.3\% showing a significantly large effect, and reduces SPD, AOD, and EOD bias by 46.9\%, 58.1\%, and 51.0\%, respectively. This effectiveness surpasses the state-of-the-art by five percentage points in the proportion of fairness improvement scenarios, and by an average of three percentage points in reducing SPD, AOD, and EOD biases.\color{black}}

\begin{figure}[!h]
    \centering
    \includegraphics[width=1\linewidth]{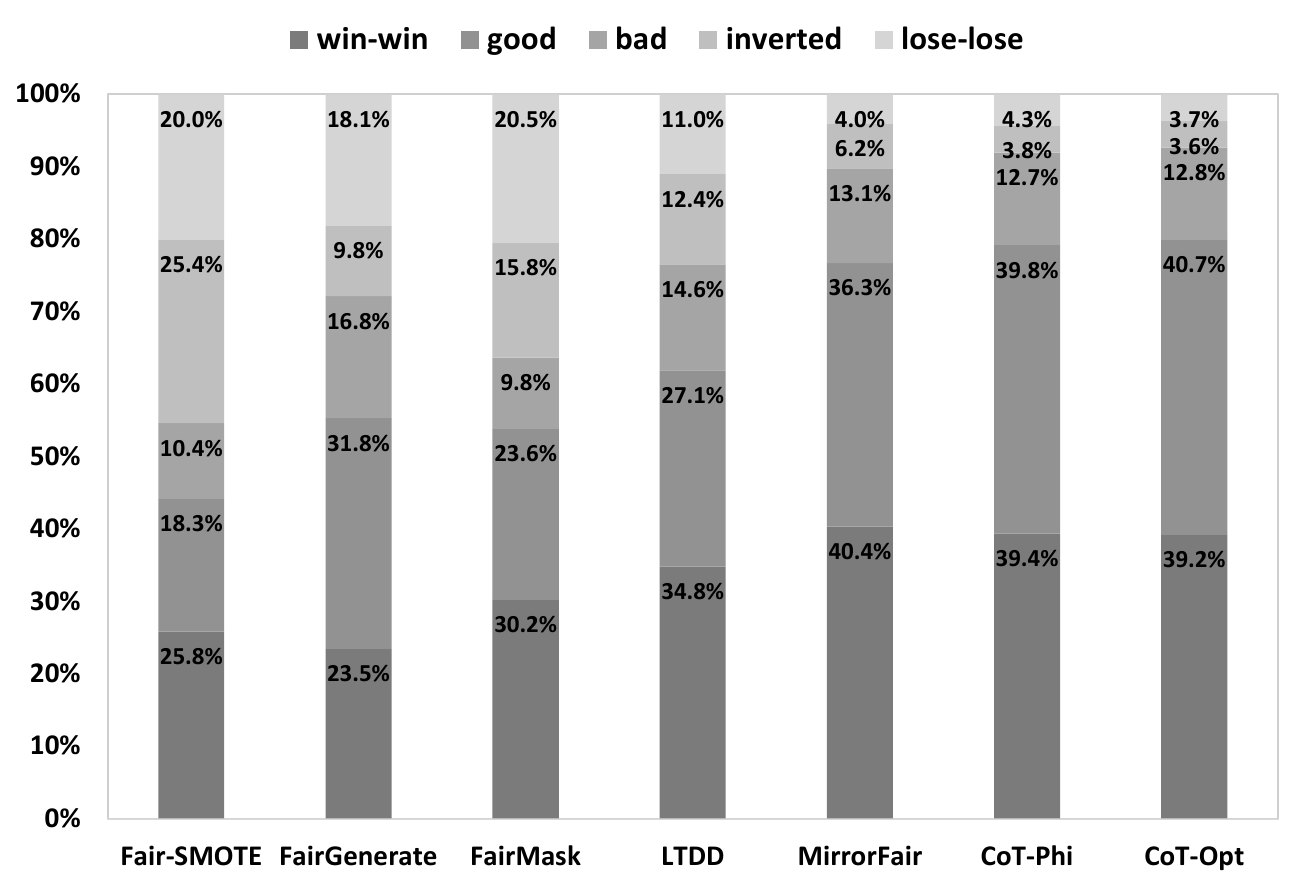}
    \caption{(RQ3) Effectiveness level distribution of existing methods and CoT in the performance-fairness trade-off. Both CoT-Phi and CoT-Opt outperform state-of-the-art methods in balancing model performance and fairness. \color{black}Specifically, CoT-Phi surpasses the Fairea-trade-off baseline in 79.2\% of cases (win-win + good), and CoT-Opt achieves this in 79.9\%, while the state-of-the-art achieves this in 76.7\% of cases.\color{black}}
    \label{fig:fairea_single}
\end{figure}

\subsection{RQ3: Performance-Fairness Trade-off}
\label{RQ3}

This RQ explores the effectiveness of balancing model performance and fairness across existing methods and CoT approaches. We employ hybrid evaluation criteria and metrics, building upon previous research. To ensure consistency with prior work \cite{hort2021fairea, chen2022maat, chen2023comprehensive}, we utilize the Fairea-Baseline to assess the trade-off effectiveness between performance and fairness for each bias mitigating method. \color{black}The evaluation includes 15 performance-fairness baselines (5 performance metrics $\times$ 3 fairness metrics) and 60 decision-making scenarios across ten decision-making tasks and six models\color{black}.

Figure \ref{fig:fairea_single} illustrates the effectiveness distribution for existing methods and CoT. Both implementations of CoT outperform the state-of-the-art in exceeding the Fairea trade-off baseline (located in the ``win-win'' and ``good'' regions). \color{black}Specifically, CoT-Phi surpasses the trade-off baseline in 79.2\% of cases, while CoT-Opt achieves this in 79.9\%, outperforming the state-of-the-art method MirrorFair by three percentage points. Additionally, as methods of the same type, both CoT-Phi and CoT-Opt outperform the state-of-the-art pre-processing approach LTDD (61.9\% ``win-win'' + ``good'' proportion) by more than 18 percentage points\color{black}.

\finding{Both CoT-Phi and CoT-Opt outperform the state-of-the-art in the model performance-fairness trade-off. \color{black}Specifically, CoT-Opt surpasses the Fairea Trade-off Baseline in 79.9\% of cases, outperforming the top post-processing method MirrorFair by three percentage points, and exceeding the leading pre-processing method LTDD by 18 percentage points.\color{black}}

% FITNESS surpasses the Fairea Baseline in 91\% (“win-win'' + ``good'') of cases, while the state-of-the-art method MAAT achieves such proportion by 88\%. 

\subsection{RQ4: Multiple Attributes Bias Mitigation}
\label{RQ4}
This RQ explores the impact of fairness improvement methods on the bias of unconsidered sensitive attributes when mitigating the bias of considered sensitive attributes, and evaluates the effectiveness of existing methods and CoT in addressing intersectional bias related to multiple sensitive attributes.

\begin{table}[htbp]
  \centering
  \caption{(RQ4) Relative changes in bias metrics of considered and unconsidered sensitive attributes achieved by existing methods and CoT. The top three values for the considered attribute (SPD1, AOD1, EOD1) are highlighted in \textbf{bold}, while the bottom three absolute effect for the unconsidered attribute (SPD2, AOD2, EOD2) are highlighted with \colorbox[rgb]{.651, .651, .651}{gray background}. LTDD, FairMask, MirrorFair, CoT-Phi, and CoT-Opt exhibit much less negative side effects (no more than 5\%) on the bias of unconsidered attributes compared to Fair-SMOTE and FairGenerate when mitigating the bias of a considered attribute.}
  \resizebox{\linewidth}{!}{

% Table generated by Excel2LaTeX from sheet 'Unconsidered_Attr'
% Table generated by Excel2LaTeX from sheet 'Unconsidered_Attr'
% Table generated by Excel2LaTeX from sheet 'Unconsidered_Attr'
% Table generated by Excel2LaTeX from sheet 'RQ4-1'
\begin{tabular}{l|rrr|rrr}
\hline
\multirow{2}[4]{*}{Method} & \multicolumn{3}{c|}{Considered Attribute} & \multicolumn{3}{c}{Unconsidered Attribute} \bigstrut\\
\cline{2-7}      & SPD1  & AOD1  & EOD1  & SPD2  & AOD2  & EOD2 \bigstrut\\
\hline
Fair-SMOTE & 3.6\% & -12.9\% & -29.3\% & 97.2\% & 92.7\% & 49.6\% \bigstrut[t]\\
FairGenerate & -10.6\% & -32.9\% & \textbf{-43.4\%} & 110.5\% & 95.5\% & 37.5\% \\
LTDD  & -29.2\% & -9.8\% & -3.1\% & -1.5\% & \cellcolor[rgb]{ .651,  .651,  .651}-0.2\% & \cellcolor[rgb]{ .651,  .651,  .651}-0.9\% \\
FairMask & -24.4\% & -41.2\% & -41.3\% & \cellcolor[rgb]{ .651,  .651,  .651}-0.8\% & 3.5\% & 2.6\% \\
MirrorFair & \textbf{-50.9\%} & \textbf{-51.9\%} & \textbf{-42.0\%} & -1.5\% & -1.0\% & -3.9\% \\
CoT-Phi & \textbf{-57.4\%} & \textbf{-54.6\%} & -40.6\% & \cellcolor[rgb]{ .651,  .651,  .651}-0.9\% & \cellcolor[rgb]{ .651,  .651,  .651}0.1\% & \cellcolor[rgb]{ .651,  .651,  .651}-0.2\% \\
CoT-Opt & \textbf{-46.9\%} & \textbf{-58.1\%} & \textbf{-51.0\%} & \cellcolor[rgb]{ .651,  .651,  .651}-0.4\% & \cellcolor[rgb]{ .651,  .651,  .651}0.3\% & \cellcolor[rgb]{ .651,  .651,  .651}0.1\% \bigstrut[b]\\
\hline
\end{tabular}%

    }
  \label{tab:unconsidered_attr}%
\end{table}%

The concepts of considered and unconsidered sensitive attributes are defined based on the objective of applying fairness improvement methods. For example, in the Adult dataset, ``sex'' and ``race'' are typically regarded as sensitive attributes requiring protection. When the target is to mitigate sex-related bias, ``sex'' is identified as the considered attribute and ``race'' as the unconsidered attribute, and vice versa. Table~\ref{tab:unconsidered_attr} presents the relative changes in bias metrics of considered and unconsidered sensitive attributes achieved by existing methods and CoT. Overall, LTDD, FairMask, MirrorFair, CoT-Phi, and CoT-Opt exhibit much less negative side effects (no more than 5\%) on the bias of unconsidered attributes compared to Fair-SMOTE and FairGenerate when mitigating the bias of a considered attribute. Notably, CoT-Opt achieves the highest overall bias reduction on the considered attribute while simultaneously achieving the lowest negative effect on the unconsidered attribute.

The initial LTDD and FairGenerate methods are not applicable to multiple sensitive attributes, and we have not identified an appropriate strategy to generalize them to a multi-attribute context. Therefore, we exclude LTDD and FairGenerate from this comparison. Table \ref{tab:intersentional_change} presents the absolute and relative changes (in parentheses) in intersectional bias metrics achieved by three existing methods and CoT. \color{black}CoT-Opt maintains a strong advantage in improving overall intersectional fairness, reducing ISPD, IAOD, and IEOD by 38.5\%, 42.2\%, and 34.0\%, respectively. Additionally, CoT-Opt achieves intersectional fairness improvement in 97.8\% of scenarios, with 74.4\% of those showing a significantly large effect. 

Regarding CoT-Phi, it achieves the largest reduction in ISPD at 50.3\%, surpassing MirrorFair by five percentage points and outperforming FairMask by 29 percentage points. CoT-Phi also attains a 83.3\% proportion of scenarios with significantly large improvements in intersectional fairness, exceeding the leading method FairMask by 13 percentage points\color{black}.

% Table generated by Excel2LaTeX from sheet 'Intersectional'
\begin{table}[htbp]
  \centering
  \caption{(RQ4) Absolute and relative changes (in parentheses) in intersectional bias metrics achieved by existing methods and CoT. The top three values are highlighted with \colorbox[rgb]{.651, .651, .651}{gray background}. \color{black}CoT-Phi achieves the best overall effectiveness, reducing ISPD, IAOD, and IEOD by 50.3\%, 42.1\%, and 29.5\%, respectively. Additionally, CoT-Opt achieves 94.4\% proportion of intersectional fairness improved scenarios and 83.2\% of them with a significantly large effect.\color{black}}
  \resizebox{\linewidth}{!}{
% Table generated by Excel2LaTeX from sheet 'Intersectional'
% Table generated by Excel2LaTeX from sheet 'Intersectional'
\begin{tabular}{l|rr|rr|rr|r|r}
\hline
\multicolumn{1}{c|}{\multirow{2}[2]{*}{Method}} & \multicolumn{2}{c|}{\multirow{2}[2]{*}{ISPD}} & \multicolumn{2}{c|}{\multirow{2}[2]{*}{IAOD}} & \multicolumn{2}{c|}{\multirow{2}[2]{*}{IEOD}} & \multicolumn{1}{r|}{\multirow{2}{*}{%
  \shortstack{Intersectional\\Fairness $\uparrow$}}} & \multicolumn{1}{r}{\multirow{2}{*}{%
  \shortstack{Significantly\\Large Effect}}} \bigstrut[t]\\
      & \multicolumn{2}{c|}{} & \multicolumn{2}{c|}{} & \multicolumn{2}{c|}{} &       &  \bigstrut[b]\\
\hline
Fair-SMOTE & -0.012 & (-3.20)\% & -0.029 & (-11.2\%) & -0.041 & (-17.4\%) & 67.8\% & 47.8\% \bigstrut[t]\\
FairMask & -0.034 & (-21.3\%) & \cellcolor[rgb]{ .651,  .651,  .651}-0.049 & \cellcolor[rgb]{ .651,  .651,  .651}(-34.4\%) & \cellcolor[rgb]{ .651,  .651,  .651}-0.067 & \cellcolor[rgb]{ .651,  .651,  .651}(-32.3\%) & \cellcolor[rgb]{ .651,  .651,  .651}95.6\% & \cellcolor[rgb]{ .651,  .651,  .651}70.0\% \\
MirrorFair & \cellcolor[rgb]{ .651,  .651,  .651}-0.069 & \cellcolor[rgb]{ .651,  .651,  .651}(-44.8\%) & -0.04 & (-26.1\%) & -0.046 & (-19.0\%) & 86.7\% & 66.7\% \\
CoT-Phi & \cellcolor[rgb]{ .651,  .651,  .651}\textbf{-0.075} & \cellcolor[rgb]{ .651,  .651,  .651}\textbf{(-50.3\%)} & \cellcolor[rgb]{ .651,  .651,  .651}\textbf{-0.063} & \cellcolor[rgb]{ .651,  .651,  .651}(-42.1\%) & \cellcolor[rgb]{ .651,  .651,  .651}-0.064 & \cellcolor[rgb]{ .651,  .651,  .651}(-29.5\%) & \cellcolor[rgb]{ .651,  .651,  .651}94.4\% & \cellcolor[rgb]{ .651,  .651,  .651}\textbf{83.3\%} \\
CoT-Opt & \cellcolor[rgb]{ .651,  .651,  .651}-0.063 & \cellcolor[rgb]{ .651,  .651,  .651}(-38.5\%) & \cellcolor[rgb]{ .651,  .651,  .651}-0.062 & \cellcolor[rgb]{ .651,  .651,  .651}\textbf{(-42.2\%)} & \cellcolor[rgb]{ .651,  .651,  .651}\textbf{-0.07} & \cellcolor[rgb]{ .651,  .651,  .651}\textbf{(-34.0\%)} & \cellcolor[rgb]{ .651,  .651,  .651}\textbf{97.8\%} & \cellcolor[rgb]{ .651,  .651,  .651}74.4\% \bigstrut[b]\\
\hline
\end{tabular}%

    }
  \label{tab:intersentional_change}%
\end{table}%

\finding{Both CoT-Phi and CoT-Opt achieve the top overall bias reduction on the considered attribute while simultaneously exhibiting the lowest negative effect on the unconsidered attribute. \color{black}In terms of intersectional fairness improvement, CoT-Opt reduces ISPD, IAOD, and IEOD by 38.5\%, 42.2\%, and 34.0\%, respectively, and improves intersectional fairness in 97.8\% of scenarios, with 74.4\% of those showing a significantly large effect. CoT-Phi achieves the largest reduction in ISPD at 50.3\%, surpassing the state-of-the-art by five percentage points, and attains a 83.3\% proportion of scenarios with significantly large improvements in intersectional fairness, exceeding the state-of-the-art by 13 percentage points.\color{black}}

% CoT-Opt maintains a strong advantage in improving overall intersectional fairness, reducing ISPD, IAOD, and IEOD by 40.0\%, 40.9\%, and 32.1\%, respectively. Additionally, CoT-Opt achieves intersectional fairness improvement in 96.7\% of scenarios, with 71.7\% of those showing a significantly large effect. 

% Regarding CoT-Phi, it achieves the largest reduction in ISPD at 51.6\%, surpassing MirrorFair by 10 percentage points and outperforming FairMask by 19 percentage points. CoT-Phi also attains a 76.7\% proportion of scenarios with significantly large improvements in intersectional fairness, exceeding the leading method FairMask by eight percentage points.

\subsection{RQ5: Robustness of CoT}
\label{RQ5}

% To investigates the robustness of CoT under overfitting and realistic data situation such as missing values, noise, and outliers, this RQ manually creating 

% two sub research questions 

This RQ investigates the robustness of CoT under overfitting and realistic data conditions using the modern classifier XGBoost across all ten tasks we studied. For the overfitting scenario, we randomly sample 100 data instances from the original training data for model training to deliberately construct an overfitting setting. For realistic data conditions, we manually contaminate it with missing values, noise, and outliers. Table \ref{tab:OverfittingContamination} presents the relative changes in model performance and fairness metrics under overfitting and realistic data conditions. Both overfitting and realistic data conditions consistently reduce model performance. The decreases across the five performance metrics range from 4 to 28 percentage points. Regarding fairness, the AOD and EOD metrics increase, while the SPD metric decreases under both conditions.

% and increase AOD and EOD.

% Table generated by Excel2LaTeX from sheet 'RQ5'
\begin{table}[htbp]
  \centering
  \caption{(RQ5) Relative change in model performance and fairness under overfitting and realistic data conditions. Overfitting and realistic data conditions can significantly reduce the original model's performance but have varying impacts on different fairness metrics.}
  \resizebox{\linewidth}{!}{
    \begin{tabular}{l|rrrrr|rrr}
    \hline
    \multirow{2}[4]{*}{Scenario} & \multicolumn{5}{c|}{Performance (+)}  & \multicolumn{3}{c}{Fairness (-)} \bigstrut\\
\cline{2-9}          & Accuracy & Recall & Precision & F1    & MCC   & SPD   & AOD   & EOD \bigstrut\\
    \hline
    Overfitting & -4\%  & -5\%  & -7\%  & -6\%  & -21\% & -10\% & 8\%   & 12\% \bigstrut[t]\\
    Realistic Data Conditions			 & -4\%  & -10\% & -5\%  & -11\% & -28\% & -20\% & 10\%  & 20\% \bigstrut[b]\\
    \hline
    \end{tabular}%
    }
  \label{tab:OverfittingContamination}%
\end{table}%

Table \ref{RQ5} presents the relative changes in accuracy and three bias metrics of the XGB model under overfitting and realistic data conditions after applying the CoT approaches. Both CoT-Phi and CoT-Opt consistently improve model fairness without reducing model accuracy. Specifically, CoT-Phi reduces SPD, AOD, and EOD by 41\%, 34\%, and 24\% under the overfitting scenario, and by 50\%, 54\%, and 52\% under realistic data conditions.

% The overfitting and realistic data conditions can significant impact the model accuracy. Specifically, under the overfitting scenario, the accuracy of default model decrease from 0.85 to 0.76, the bias remains significant. However, both CoT-Phi and CoT-Opt maintains the effectiveness and reduce the SPD, AOD, and EOD bias from 0.17, 0.16, and 0.22 to 0.08/0.08, 0.06/0.06, and 0.09/0.08, respectively.

% Regarding, realistic data condition, the accuracy of default model decrease from 0.85 to 0.80, and bias remains. CoT-Phi and CoT-Opt reduce the SPD, AOD, and EOD bias from 0.13, 0.16, and 0.27 to 0.03/0.04, 0.02/0.02, and 0.02/0.03, respectively.

% Table generated by Excel2LaTeX from sheet 'Intersectional'
\begin{table}[htbp]
  \centering
  \caption{(RQ5) The relative changes in model accuracy, SPD, AOD, and EOD after applying CoT under overfitting and realistic data conditions. Specifically, CoT consistently improves model fairness without reducing model accuracy under both overfitting and realistic data conditions.}
  \resizebox{\linewidth}{!}{
% Table generated by Excel2LaTeX from sheet 'Intersectional'
% Table generated by Excel2LaTeX from sheet 'Intersectional'
% Table generated by Excel2LaTeX from sheet 'RQ5'
% Table generated by Excel2LaTeX from sheet 'RQ5'
\begin{tabular}{l|rrrr|rrrr}
\hline
\multirow{2}[4]{*}{Method} & \multicolumn{4}{c|}{Overfitting} & \multicolumn{4}{c}{Realistic Data Conditions} \bigstrut\\
\cline{2-9}      & Accuracy & SPD   & AOD   & EOD   & Accuracy & SPD   & AOD   & EOD \bigstrut\\
\hline
CoT-Phi & 0\%   & -41\% & -34\% & -24\% & 0\%   & -50\% & -54\% & -52\% \bigstrut[t]\\
CoT-Opt & 0\%   & -35\% & -32\% & -25\% & 0\%   & -35\% & -41\% & -41\% \bigstrut[b]\\
\hline
\end{tabular}%

    }
  \label{tab:robustness}%
\end{table}%

\color{black}

% \finding{\color{black}Both CoT-Phi and CoT-Opt are robustly mitigating bias in overfitting and realistic data conditions. Specifically, CoT-Opt  reduces the SPD, AOD, EOD from 0.17/0.13, 0.16/0.16, and 0.22/0.27, to 0.08/0.04, 0.06/0.02, and 0.08/0.03, respectively.\color{black}}

\finding{\color{black} CoT are robust in repairing fairness issues under overfitting and realistic data conditions. Specifically, CoT-Phi reduces SPD, AOD, and EOD by 41\%/50\%, 34\%/54\%, and 24\%/52\% in overfitting and realistic data conditions, respectively.\color{black}}

\section{Implication and Discussion}
\label{sec:discussion}

In this section, we discuss the implications and advantages of CoT, as well as the threats to the validity of our experimental results.

\subsection{Analysis of CoT}

\subsubsection{Effectiveness}
CoT outperforms baseline methods across all measurements, including absolute and relative changes in fairness metrics, the proportion of fairness improvement scenarios, and the fairness-performance trade-off, in both single and multiple sensitive attribute scenarios, across ten tasks and four classification algorithms. This extensive evaluation gives us confidence to claim the effectiveness of CoT.

\subsubsection{Applicability}
As a pre-processing method, CoT focuses on data debugging and is model-agnostic, making it applicable to any AI model. It overcomes the limitations of current state-of-the-art methods such as MirrorFair, which rely on post-prediction probability tuning and are therefore difficult to apply to models that do not explicitly output class probabilities~\cite{xiao2024mirrorfair}.

\subsubsection{Efficiency}

Table \ref{tab:ExecutionTime} presents the execution time of the bias mitigation methods. LTDD and CoT-Phi do not require any additional model training, resulting in almost no increased overhead. MirrorFair and FairMask require training a mirror model and an extrapolation model, respectively, which effectively doubles the model training overhead. CoT-Opt increases computational cost due to its optimization process. Fair-SMOTE and FairGenerate synthesize new data instances, leading to computational overhead that is dozens of times higher. CoT-Phi achieves state-of-the-art effectiveness without increasing computational cost, making it the most efficient approach.

\subsubsection{Feasibility}

Both CoT-Phi and CoT-Opt only require processing the training data, without the need for additional adjustments or extra components in the software deployment pipeline. This demonstrates greater feasibility compared to current state-of-the-art methods such as MirrorFair, which requires deploying an additional mirror model.

% Table generated by Excel2LaTeX from sheet 'Time_overhead'
\begin{table}[t]
  \centering
  \caption{Execution time (in seconds) for existing methods and CoT over 20 runs across four tasks using the LR model.}
  \resizebox{\linewidth}{!}{
% Table generated by Excel2LaTeX from sheet 'Time_overhead'
\begin{tabular}{l|rrrr}
\hline
Method & Adult-Sex & Compas-Sex & MEP1-Sex  & Default-Sex \bigstrut\\
\hline
Fair-SMOTE & 196   & 28    & 130   & 345 \bigstrut[t]\\
FairGenerate & 143   & 19    & 105   & 273 \\
CoT-Opt & 22    & 8     & 20    & 19 \\
FairMask & 8     & 3     & 11    & 18 \\
MirrorFair & 6     & 3     & 6     & 5 \\
LTDD  & 4     & 3     & 4     & 4 \\
CoT-Phi & 4     & 3     & 4     & 4 \\
Original Model & 4     & 3     & 4     & 4 \bigstrut[b]\\
\hline
\end{tabular}%
}

  \label{tab:ExecutionTime}%
\end{table}%

\color{black}
\subsubsection{Scalability}

Scalability is an important aspect of any bias mitigation approach. CoT mitigates bias by processing sensitive attributes, meaning that increases in the dimensionality of non-sensitive features have minimal impact. In terms of data size and the number of sensitive attributes, the computational overhead scales linearly. Regarding effectiveness scalability, Table \ref{tab:unconsidered_attr} shows that CoT has limited effect on unconsidered sensitive attributes. Therefore, CoT demonstrates good scalability with respect to data size, feature dimensionality, and the number of sensitive attributes. Since CoT operates at the data level and is model-agnostic, we did not observe any inherent scalability issues. Notably, when CoT is applied in conjunction with language models on tabular tasks, the data must first be converted into prompt–response pairs. In such cases, very high-dimensional feature spaces may approach or exceed the input length limits of current language models, which could constrain the applicability of CoT.

% \subsubsection{Scalability}

% Scalability is significant for a bias mitigation approach. CoT mitigates bias via processing the sensitive attributes so increases in the dimensionality of non-sensitive features have little effect. Regarding the data size and the number of sensitive attributes the computational overhead scale linearly. Regarding the effectiveness scalability, Table \ref{tab:unconsidered_attr} demonstrates CoT have less effect on unconsedered sensitive attribute. Therefore, CoT have good scalability in data size, feature size, and sensitive attribute set size level. Because CoT is applied at the data level and is model-agnostic, we are not aware of inherent scalability issues. That said, when CoT is used in conjunction with LLMs on tabular tasks, the data must first be transformed into prompt–response pairs. In such cases, very high-dimensional feature spaces may approach or exceed the input length limits of current LLMs, which could constrain the applicability of CoT. We will add a discussion of this practical limitation in the revised manuscript.

\subsubsection{Reflecting the Real-world Distributions}
% Unlike data augmentation techniques such as Fair-SMOTE \cite{chakraborty2021bias}, which generate entirely virtual data instances, CoT directly modifies the sensitive attributes while keeping all other features and data labels unchanged. As a result, the transformed data continues to closely reflect real-world distributions. By preserving the joint distribution of non-sensitive features and labels, CoT effectively preserve the underlying decision logic of original data. This alignment is further supported by the comparison of model performance before and after applying CoT. Nevertheless, we encourage practitioners to restrict the use of CoT-processed data solely to model training, rather than for other purposes.

Unlike data augmentation techniques such as Fair-SMOTE \cite{chakraborty2021bias}, which generate entirely virtual data instances, CoT directly modifies the sensitive attributes while keeping all other features and data labels unchanged. As a result, the transformed data continues to closely reflect real-world distributions. This alignment is further supported by the comparison of model performance before and after applying CoT. Nevertheless, we encourage practitioners to restrict the use of CoT-processed data solely to model training, rather than for other purposes.

\color{black}

\subsection{Impact of Optimization Strategy Design}
In Section \ref{sec:non-sensitive_attribute_bias}, we introduce our default implementation of CoT-Opt for achieving optimal overall bias mitigation effectiveness. In practice, CoT-Opt can be tailored to prioritize specific fairness metrics by modifying the loss function. For example, removing SPD and AOD from the loss function while retaining EOD can enhance effectiveness in reducing EOD bias. Regarding the optimization algorithm, we use Particle Swarm Optimization (PSO) \cite{eberhart1995particle} to accelerate the optimization process. Other algorithms, such as Genetic Algorithms (GA) \cite{whitley1994genetic}, are also applicable. The choice of optimization algorithm does not impact the effectiveness of CoT-Opt, but it can influence optimization speed. Due to page limitations, we are unable to present all of our experimental results for different optimization strategies in this paper. However, the related files are available in our supplementary material.

\subsection{Threats to Validity}

\subsubsection{Internal Validity}
\label{sec:internal_validity}
Potential threats to the internal validity of our results stem from the selection of benchmark datasets, tasks, evaluation criteria, and baseline methods. Many bias mitigation techniques are evaluated using different benchmark datasets and fairness metrics in their original papers, complicating direct comparisons between CoT and baseline approaches and potentially impacting the validity of our experiments. To address these concerns, we align with prior research \cite{chakraborty2020fairway, chakraborty2021bias, chen2022maat, peng2022fairmask, chen2024fairness}, adopting widely recognized benchmark datasets and employing multiple evaluation metrics. Nevertheless, it should be noted that results may differ slightly from those reported in original studies if new datasets, algorithms, or metrics are introduced.

% \subsubsection{External Validity}
% CoT is specifically designed to address bias mitigation in tabular data and classification tasks, which are prevalent in software applications. However, the generalizability of CoT to alternative scenarios, such as natural language processing or computer vision, requires further investigation.

\subsubsection{External Validity} CoT is specifically designed to address bias mitigation in tabular data and classification tasks, which are prevalent in software applications. However, several factors may limit its external validity. While we evaluated CoT across ten diverse benchmark tasks, these datasets primarily represent structured tabular data. The effectiveness and generalizability of CoT to alternative scenarios with unstructured data, such as natural language processing (NLP) or computer vision (CV), require further investigation. For instance, extracting and sensitive attribute and data labels from high-dimensional image pixels or complex linguistic structures may present unique challenges. Furthermore, our study mainly considered binary sensitive attributes. In real-world software deployments, sensitive attributes can be non-binary or continuous and the applicability of the Phi-coefficient in these continuous correlation contexts warrants additional empirical validation.

% \subsubsection{Internal Validity}
% \label{sec:internal_validity}
% The potential threats to the internal validity of our results arise from the selection of benchmark datasets, tasks, evaluation criteria, and existing methods. 
% Many bias-mitigating methods are evaluated using different benchmark datasets as well as different fairness metrics in their original papers, which increases the complexity of the comparisons between CoT and baseline methods and can threaten the validity of our experiments.
% To mitigate these potential concerns, we align with prior research \cite{chakraborty2020fairway, chakraborty2021bias, chen2022maat, peng2022fairmask, chen2024fairness}, opting for widely recognized benchmark datasets and employing multiple evaluation metrics. 
% However, it is noteworthy that the outcomes might exhibit slight variations compared with their original papers if new datasets, algorithms, and evaluation metrics are introduced. 

% \subsubsection{External Validity}
% CoT has been meticulously crafted to address bias mitigation within tabular data and classification tasks, both of which are prevalent in the realm of software. However, the validity of CoT in alternative scenarios, such as natural language processing (NLP) or computer vision (CV), necessitates a thorough investigation.

\color{black}

% \subsection{Exploration of CoT in TinyLlama}

% Recently, language models have demonstrated promising capabilities across various domains \cite{brown2020language, bommasani2021opportunities}, including tabular data tasks \cite{zhang2024tablellm}. Thereby, we have conducted a preliminary exploration of the effectiveness of CoT in language models for tabular data tasks. 

% We adopted TinyLlama-1.1B, a compact language model from Meta \cite{zhang2024tinyllama}, to conduct our exploration. Our experimental results show that TinyLlama achieves performance comparable to classic ML models (e.g., LR and XGB) and demonstrates similar SPD, AOD, and EOD biases, on low-dimensional datasets (fewer than 30 features), such as Adult, COMPAS, and Default. CoT maintains its effectiveness within TinyLlama for these tasks. However, on high-dimensional datasets (more than 40 features), such as MEP1 and MEP2, TinyLlama fails, with accuracies dropping below 0.50. Under such circumstances, it is not appropriate to claim the effectiveness of CoT by comparing performance and fairness metrics before and after its application on TinyLlama for these datasets. Therefore, our preliminary findings indicate that while language models can be applied to tabular data tasks and CoT remains effective, the feature dimensionality significantly impacts model performance. We next introduce our future work based on our preliminary results.

% For better logical flow and readability, we present these results and implementation details in the supplementary material rather than in the main paper.

\subsection{Future Work}

Recently, large language models (LLMs) have demonstrated promising capabilities across various domains \cite{brown2020language, bommasani2021opportunities, zhang2024tablellm}. Therefore, we plan to explore extending CoT to typical LLM tasks such as text classification and reasoning. Since CoT tunes correlations within training data based on explicitly defined sensitive attributes and labels, it is feasible to extend CoT to text classification tasks by extracting sensitive attributes and labels directly from the natural language text. We plan to pursue this direction in our future work.

\color{black}

\section{Conclusion}
\label{sec:conclusion}

This paper investigates software fairness, emphasizing that fairness concerns are not only ethical but also constitute a significant software quality issue stemming from software performance bugs. We further highlight the practical significance of software fairness research, as bias mitigation methods empower software systems to dynamically adjust performance across groups, enhancing both out-of-distribution generalization and geographical transferability. We introduce CoT, an effective, widely applicable, engineering-friendly, and efficient pre-processing bias mitigation approach based on the Phi-coefficient and multi-objective optimization. CoT demonstrates clear advantages over current state-of-the-art methods. Additionally, the successful application of the Phi-coefficient and multi-objective optimization in CoT opens up new research opportunities for the AI and SE communities, including balancing the software robustness, efficiency, privacy, trustworthiness, and performance.

\newpage

\bibliographystyle{ACM-Reference-Format}
\bibliography{0-references}

\end{document}